\documentclass[aps,prx,superscriptaddress,amsfonts,amsmath,amssymb,showpacs,floatfix,reprint]{revtex4-1}

\usepackage{bm}
\usepackage{graphicx}
\usepackage{amsmath}
\usepackage{amstext}
\usepackage{amssymb}
\usepackage{amsfonts}
\usepackage{amsbsy}
\usepackage{verbatim}
\usepackage[usenames,dvipsnames,svgnames,table]{xcolor}
\usepackage{multirow}
\usepackage{gensymb}
\usepackage{textcomp}
\usepackage{enumitem}
\usepackage{fancyhdr}
\usepackage[colorlinks=true, urlcolor=Blue, linkcolor=Blue, citecolor=Blue, pdftex]{hyperref}
\usepackage{times}
\usepackage{floatrow}
\usepackage{dcolumn}
\usepackage{bigdelim}
\usepackage{siunitx}

\def\qqq{{(q,q,q)}}
\def\qq0{{(q,q,0)}}
\def\q00{{(q,0,0)}}

\def\tpmqq0{{(2\pi-q,q,0)}}
\def\tpmq00{{(2\pi-q,0,0)}}
\def\000{{(0,0,0)}}
\def\2p00{{(2\pi,0,0)}}

\def\pp0{{(\pi,\pi,0)}}

\def\<{\langle}
\def\>{\rangle}
\newcommand{\JMMM}{J.~Magn. Magn. Mater.}
\newcommand{\JSTAT}{J.~Stat.~Mech.}
\newcommand{\PRL}{Phys.~Rev.~Lett.}
\newcommand{\JSF}{J.~Stat.~Phys.}

\begin{document}

\title{Quantum paramagnetism and helimagnetic orders in the Heisenberg model on the body centered cubic lattice}

\author{Pratyay Ghosh}
\affiliation{Department of Physics, Indian Institute of Technology Madras, Chennai 600036, India}
\author{Tobias M\"uller}
\affiliation{Institute for Theoretical Physics and Astrophysics, Julius-Maximilian's University of W\"urzburg, Am Hubland, D-97074 W\"urzburg, Germany}
\author{Francesco Parisen Toldin}
\affiliation{Institute for Theoretical Physics and Astrophysics, Julius-Maximilian's University of W\"urzburg, Am Hubland, D-97074 W\"urzburg, Germany}
\author{Johannes Richter}
\affiliation{Institut f\"ur Physik, Otto-von-Guericke-Universit\"at Magdeburg, D-39016 Magdeburg, Germany}
\affiliation{Max Planck Institute for the Physics of Complex Systems, N\"othnitzer Stra{\ss}e 38, D-01187 Dresden, Germany}
\author{Rajesh Narayanan}
\affiliation{Department of Physics, Indian Institute of Technology Madras, Chennai 600036, India}
\author{Ronny Thomale}
\affiliation{Institute for Theoretical Physics and Astrophysics, Julius-Maximilian's University of W\"urzburg, Am Hubland, D-97074 W\"urzburg, Germany}
\author{Johannes Reuther}
\affiliation{Dahlem Center for Complex Quantum Systems and Fachbereich Physik, Freie Universit{\"a}t Berlin, D-14195 Berlin, Germany}
\affiliation{Helmholtz-Zentrum f\"{u}r Materialien und Energie, Hahn-Meitner-Platz 1, 14019 Berlin, Germany} 
\author{Yasir Iqbal}
\email[]{yiqbal@physics.iitm.ac.in}
\affiliation{Department of Physics, Indian Institute of Technology Madras, Chennai 600036, India}

\date{\today}

\begin{abstract}
We investigate the spin $S=1/2$ Heisenberg model on the body centered cubic lattice in the presence of ferromagnetic and antiferromagnetic nearest-neighbor $J_{1}$, second-neighbor $J_{2}$, and third-neighbor $J_{3}$ exchange interactions. The classical ground state phase diagram obtained by a Luttinger-Tisza analysis is shown to host six different (noncollinear) helimagnetic orders in addition to ferromagnetic, N\'eel, stripe and planar antiferromagnetic orders. Employing the pseudofermion functional renormalization group (PFFRG) method for quantum spins ($S=1/2$) we find an extended nonmagnetic region, and significant shifts to the classical phase boundaries and helimagnetic pitch vectors caused by quantum fluctuations while no new long-range dipolar magnetic orders are stabilized. The nonmagnetic phase is found to disappear for $S=1$. We calculate the magnetic ordering temperatures from PFFRG and quantum Monte Carlo methods, and make comparisons to available data.  
\end{abstract}

\maketitle

\section{Introduction}\label{sec:intro}

The long-range ferromagnetic (FM) or antiferromagnetic (AF) order of spins pinned to the sites of a bipartite crystal lattice becomes frustrated in the presence of longer range AF interactions, a scenario called parametric frustration. For Heisenberg spins in the classical limit, i.e., spin $S\to\infty$, these competing interactions provide a promising route towards realizing (noncollinear) helimagnetic orders, i.e., spiral spin structures~\cite{Villain-1959,Yoshimori-1959,Nagamiya-1968,Rastelli-1979}. On the square lattice, the FM or AF ordering of spins when frustrated via AF second- and third neighbor Heisenberg interactions is known to stabilize one- and two-dimensional helimagnetic orders~\cite{Rastelli-1979}. When the reciprocal spin $1/S$ becomes nonzero, quantum fluctuations enter the picture, and their amplitude increases with increasing reciprocal spin $1/S$. In fact, the quantity $1/S$ plays the same role for quantum fluctuations as the temperature does for classical fluctuations~\cite{Kaganov-1987}, although they may act differently as has been suggested in the kagome Heisenberg AF~\cite{Harris-1992,Reimers-1993,Huse-1992,Henley-2009,Korshunov-2002,Chernyshev-2014,Goetze-2015}. In the semiclassical ($1/S \ll 1$) regime, it is known that for collinear phases the quantum corrections to the ground state and the spin wave spectrum are modest~\cite{Kaganov-1987}. On the other hand, in helimagnets, owing to the delicate interplay of competing interactions the impact of quantum fluctuations is likely to be of significance. It was shown by Chubukov~\cite{Chubukov-1984} that quantum fluctuations lead to a shift of the spiral pitch vector value, but keep the two Goldstone modes ($\mathbf{k}=\mathbf{0}$ and $\mathbf{k}=\pm\mathbf{Q}$) intact, thus preserving the general structure of the magnon spectrum. However, in the small spin\textendash$S$ limit where strong quantum fluctuations are at play, the fate of the helimagnetic ground states remains largely not studied. In particular, it is of interest to investigate whether in the extreme quantum limit of $S=1/2$, quantum fluctuations could melt the helimagnetic structures~\cite{Iqbal-2017,Iqbal-2018,Iqbal-2019a} and potentially realize a quantum paramagnetic ground state~\cite{Anderson-1973,Balents-2010,Savary2016}. In this context, low-dimensional quantum spin systems have traditionally attracted much attention due to the significant increase in the role played by quantum fluctuation effects. On the square lattice, for $S=1/2$, the helimagnetic orders give way to a quantum paramagnet over an appreciable region in parameter space~\cite{Sindzingre-2009,Sindzingre-2010,Iqbal-2016}, however, similar scenarios in three-dimensional lattices remain largely unexplored.

\begin{figure}
\includegraphics[width=0.75\columnwidth]{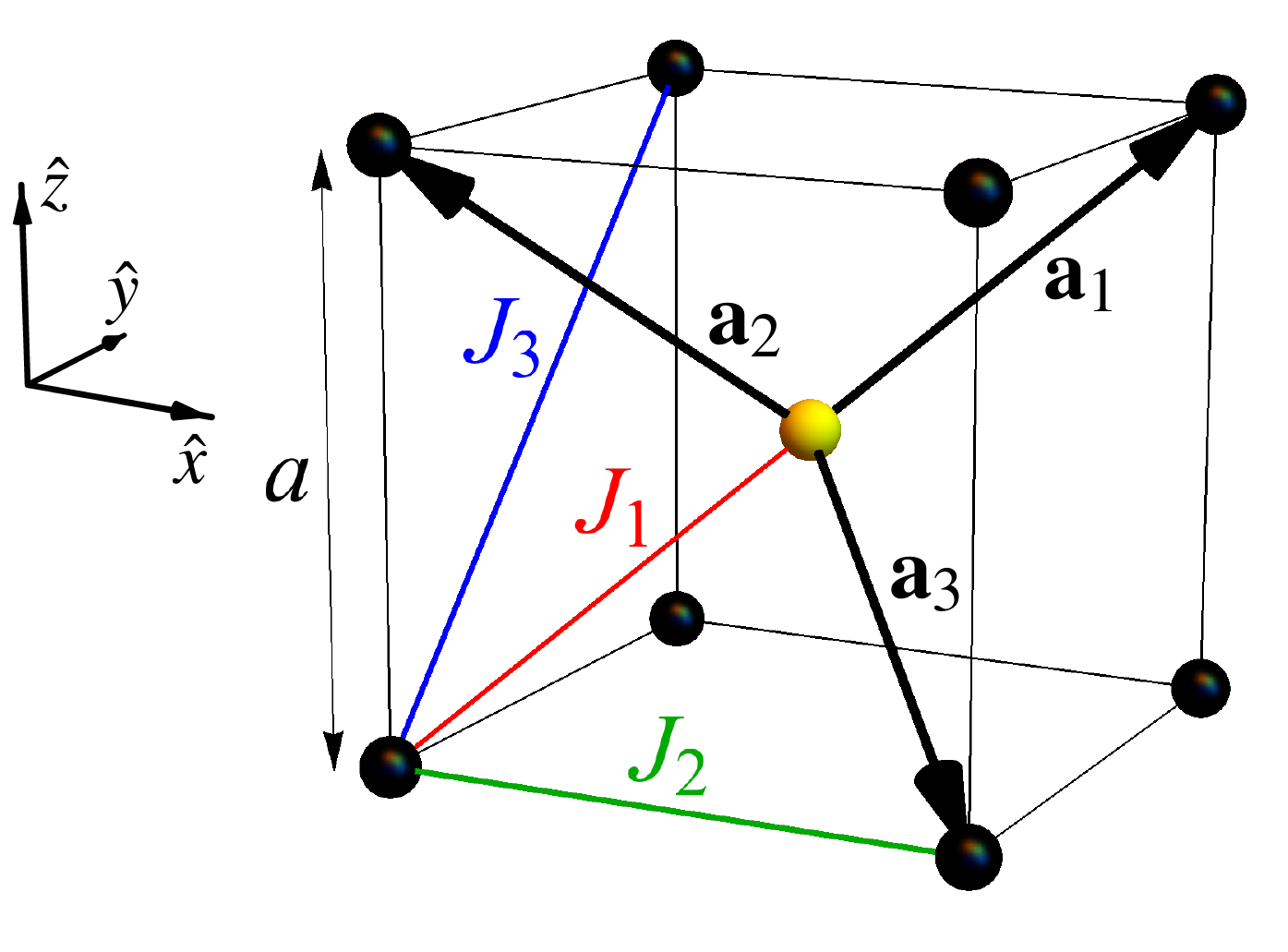}
\caption{Cubic unit cell of the BCC lattice together with its lattice vectors $\mathbf{a}_{1}=(\frac{1}{2},\frac{1}{2},\frac{1}{2})$, $\mathbf{a}_{2}=(-\frac{1}{2},-\frac{1}{2},\frac{1}{2})$, and $\mathbf{a}_{3}=(\frac{1}{2},-\frac{1}{2},-\frac{1}{2})$, assuming the lattice constant $a=1$. The Heisenberg couplings at nearest-neighbor $J_{1}$, second-neighbor $J_2$, and third-neighbor $J_3$ defining the Hamiltonian [Eq.~\eqref{eqn:heis-ham}] are also shown. The BCC lattice is a bi-bipartite lattice, i.e., it is a bipartite lattice which itself consists of two interpenetrating bipartite simple cubic lattices (labelled by black and yellow spheres) displaced by $\mathbf{a}_{1}$. The coordination numbers for different neighbors are, nearest-neighbor $z_{1}=8$, second-neighbor $z_{2}=6$, and third-neighbor $z_{3}=12$.}\label{fig:BCC-lattice}
\end{figure}

In this paper, employing the pseudofermion functional renormalization group (PFFRG) method~\cite{Reuther-2010}, we address the question as to which degree the impact of quantum fluctuations is mellowed down with an increase in dimensionality of the lattice to three spatial dimensions (3D). In order to accommodate, without frustration, both the two-sublattice N\'eel $(\mathbf{k}=(\pi,\pi))$ and the stripe $(\mathbf{k}=(\pi,0))$ orders of the square lattice in a 3D lattice, we require a bipartite lattice which itself is composed of two interpenetrating bipartite lattices, i.e., it is a bi-bipartite lattice. The body centered cubic (BCC) lattice [Fig.~\ref{fig:BCC-lattice}] has precisely this property; it is a Bravais lattice which is composed of two interpenetrating, identical simple cubic sublattices, and thus serves as a natural analogue of the square lattice in 3D~\cite{Schmidt-2002}. We investigate the classical and $S=1/2$ Heisenberg model on the BCC lattice in the presence of nearest-neighbor $J_{1}$, second-neighbor $J_{2}$ and third-neighbor $J_{3}$ exchange 
couplings,
\begin{equation}\label{eqn:heis-ham}
\hat{{\cal H}} = J_1 \sum_{\langle i,j \rangle_{1}} \mathbf{\hat{S}}_{i} \cdot \mathbf{\hat{S}}_{j}
+J_2 \sum_{\langle i,j \rangle_{2}} \mathbf{\hat{S}}_{i} \cdot \mathbf{\hat{S}}_{j}
+J_3 \sum_{\langle i,j \rangle_{3}} \mathbf{\hat{S}}_{i} \cdot \mathbf{\hat{S}}_{j},
\end{equation} 
\noindent where the $\mathbf{\hat{S}}_{i}$ are the $S=1/2$ Heisenberg spin operators on site $i$. In the classical limit ($S\to\infty$), the $\mathbf{\hat{S}}_{i}$ reduce to three-component vectors. The symbols $\langle i,j \rangle_{1}$, $\langle i,j \rangle_{2}$ and $\langle i,j \rangle_{3}$ denote sums over nearest-neighbor, second-neighbor, and third-neighbor pairs of sites, respectively. The $J_{1}$, $J_{2}$ and $J_{3}$ are allowed to be both FM and AF, and thus we will consider all possible combinations of the signs of the couplings in Eq.~\eqref{eqn:heis-ham}. Early interest and investigations into the model (with additional four spin interaction terms) stemmed from its relevance to the description of the BCC phase of solid $^{3}{\rm He}$ at low-temperatures~\cite{Utsumi-1977,Okada-1978,Yosida-1980}.   

\begin{figure*}
\includegraphics[width=0.95\columnwidth]{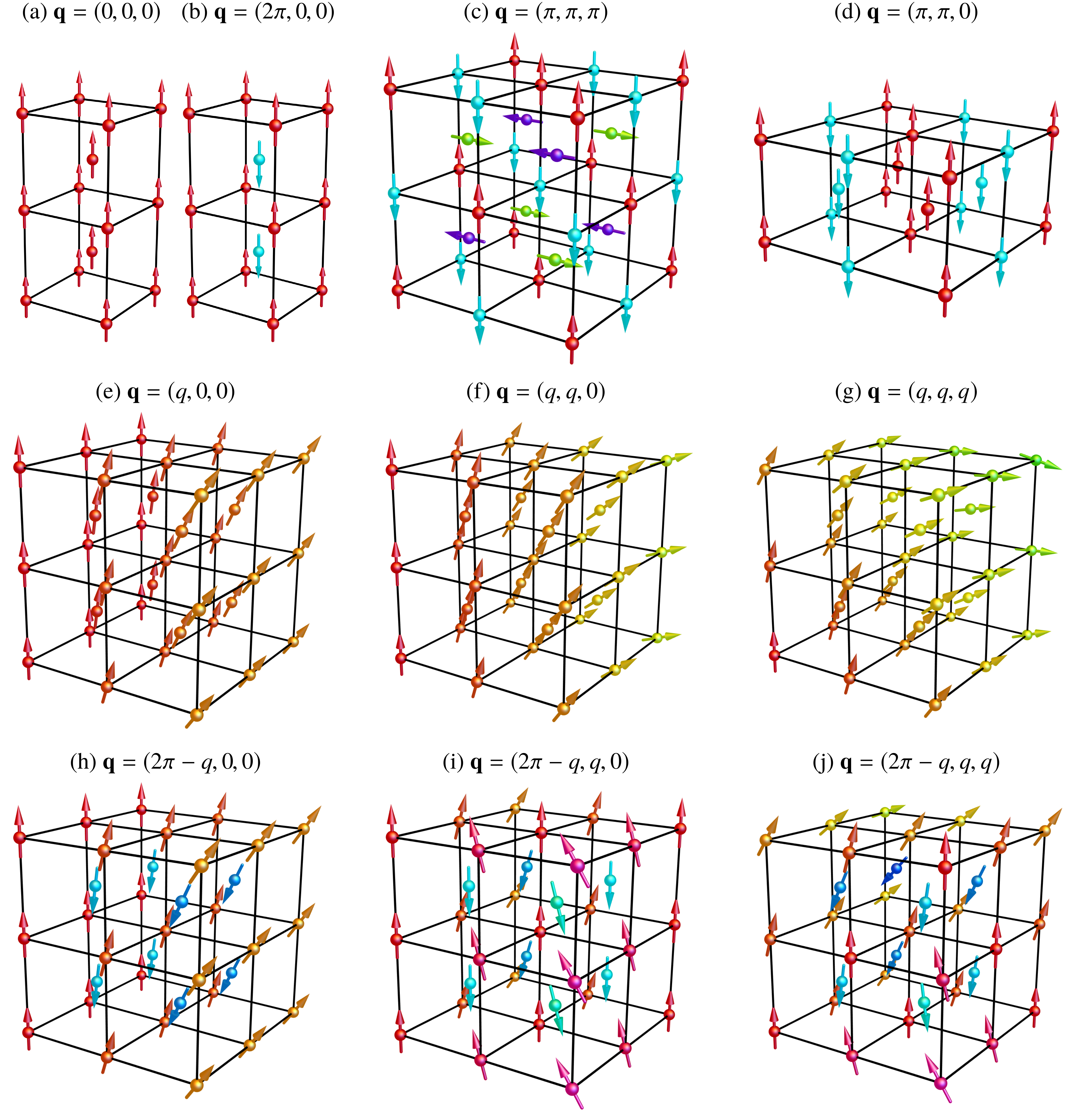}
\caption{Illustration of the classical spin configurations found in the $J_{1}$\textendash$J_{2}$\textendash$J_{3}$ Heisenberg model on the BCC lattice. The global orientation of the spins and, where applicable, the chirality of the spin spirals is not determined by model [Eq.~\eqref{eqn:heis-ham}]. The simplest ground states of this model are the (a) ferromagnet, and (b) N\'eel antiferromagnet. For the latter, the BCC lattice can be divided into two simple cubic lattices, one being the body centers of the other, which are ordered antiferromagnetically. 
(c) The $(\pi,\pi,\pi)$-state can be decomposed into two interpenetrating N\'eel-ordered simple cubic lattices, and in this illustration we choose to show them rotated by an angle $\pi/2$ relative to each other. (d) The $(\pi, \pi,0)$ antiferromagnet consists of ferromagnetic ${110}$-type planes of spins aligned antiparallel to neighboring planes. The remaining six spin configurations come in pairs of corresponding spirals for ferromagnetic and antiferromagnetic nearest neighbor coupling $J_1$. (e) In the $(q,0,0)$-state, spins are spiraling when moving along one spatial direction and are ordered ferromagnetically in the {100}-planes perpendicular to this direction. 
The corresponding $(2\pi-q,0,0)$-state (h) for antiferromagnetic $J_1$ is the same state, but with flipped spins at the bodycenter positions. (f) The $(q, q,0)$-state similarly features spins spiraling in two spatial directions with the same pitch, leading to ferromagnetically ordered {110}-planes parallel to the third direction. 
The corresponding $(2\pi-q,q,0)$-state (i) features the same spiraling behavior in two directions, but these spirals now have different chiralities, albeit the same pitch. On top of this, the spins on the body centers are flipped. Note that the relative chirality of the spirals is fixed within the Heisenberg model in contrast to the absolute chirality. The $(q,q,q)$-state (g) features spiraling with the same pitch and chirality in all three spatial directions, which means spins are ferromagnetically aligned in {111}-planes. For antiferromagnetic $J_2$, the corresponding state (j) has a wave vector of $(2\pi-q,q,q)$. This implies that the spins are spiraling backward when moving along one particular spatial direction and, in addition, the spins on the body centers are again flipped. Spins are now aligned parallel on {$-$111}-planes. \label{fig:mag-structure}}
\end{figure*}

The classical ground states of the BCC $J_{1}$\textendash$J_{2}$ model are~\cite{Utsumi-1977,Shender-1982}: (i) For $J_{2}/|J_{1}|<2/3$, a FM state [Fig.~\ref{fig:mag-structure}(a)] with $\mathbf{k}=(0,0,0)$ for FM $J_{1}$ \emph{or} a two-sublattice N\'eel state [Fig.~\ref{fig:mag-structure}(b)] with $\mathbf{k}=(2\pi,0,0)$ for AF $J_{1}$. (ii) For $J_{2}/|J_{1}|>2/3$, a stripe antiferromagnet [Fig.~\ref{fig:mag-structure}(c)] with $\mathbf{k}=(\pi,\pi,\pi)$ is stabilized in both cases, a  FM or AF $J_{1}$. This is because the transition point depends \emph{only} on the coordination number at nearest-neighbor $z_{1}$ ($=8$) and second-neighbor $z_{2}$ ($=6$) distances, with the critical $(J_{2}/|J_{1}|)_{c}=z_{1}/2z_{2}$, hence $(J_{2}/|J_{1}|)_{c}=2/3$~\cite{Utsumi-1977,Schmidt-2002}. For the corresponding $S=1/2$ BCC $J_{1}$\textendash$J_{2}$ model, all previous studies suggest a single direct phase transition from the FM or N\'eel state to the stripe ordered state~\cite{Raza-1964,Mueller-2015,Schmidt-2002,Oitmaa-2004,Majumdar-2009,Pantic-2014,Farnell-2016}. Thus, in contrast to the square lattice $S=1/2$ $J_{1}$\textendash$J_{2}$ Heisenberg model~\cite{Schulz-1992,Schulz-1996,Shannon-2006,Iqbal-2016}, 
there is an absence of an intermediate quantum paramagnetic phase, a manifestation of the weakening of quantum fluctuations in 3D. Note however, that for the square lattice model with FM $J_1$, the very existence of quantum paramagnetic phase is not very clear yet~\cite{Shindou-2011,richter2010-2d-j1j2,Iqbal-2016}. 
On the BCC lattice, the role of a further neighbor \emph{frustrating} AF $J_{3}$ coupling in Eq.~\eqref{eqn:heis-ham} has not yet been investigated, neither at the classical or semiclassical level nor in the limit of small spin\textendash$S$. At the classical level, our Luttinger-Tisza analysis shows that the inclusion of an AF $J_{3}$ coupling stabilizes a plethora of helimagnetic structures, and a planar AF order~\cite{Utsumi-1977}. In particular, for a model with FM $J_{1}$ we find three incommensurate spiral orders, namely, a 1D spiral with $\mathbf{q}=(q,0,0)$ [Fig.~\ref{fig:mag-structure}(e)], a 2D spiral with $\mathbf{q}=(q,q,0)$ [Fig.~\ref{fig:mag-structure}(f)], and a 3D spiral with $\mathbf{q}=(q,q,q)$ [Fig.~\ref{fig:mag-structure}(g)]. Similarly, in the case of AF $J_{1}$ we find three corresponding incommensurate spiral orders, namely, a 1D spiral with $\mathbf{q}=(2\pi-q,0,0)$ [Fig.~\ref{fig:mag-structure}(h)], a 2D spiral with $\mathbf{q}=(2\pi-q,q,0)$ [Fig.~\ref{fig:mag-structure}(i)], and a 3D spiral with $\mathbf{q}=(2\pi-q,q,q)$ [Fig.~\ref{fig:mag-structure}(j)]. In addition, for both FM and AF $J_{1}$, a planar AF order with $\mathbf{q}=(\pi,\pi,0)$ [Fig.~\ref{fig:mag-structure}(d)] is stabilized at large $J_{2}$ and $J_{3}$. The global classical phase diagram is presented in Fig.~\ref{fig:cpd-qpd}(a) and Fig.~\ref{fig:cpd-qpd}(c) together with the pitch vectors of these incommensurate spirals given in Table~\ref{tab:Ordering-Vec-Classical}. 

For the quantum $S=1/2$ $J_{1}$\textendash$J_{2}$\textendash$J_{3}$ model for both FM and AF $J_{1}$, our PFFRG analysis reveals that the most salient manifestation of quantum fluctuations is the realization of a quantum paramagnetic (PM) phase centered at the tricritical point of the 2D spiral, 3D spiral and planar AF orders [Fig.~\ref{fig:cpd-qpd}(b) and Fig.~\ref{fig:cpd-qpd}(d)]. This PM phase has an extended span in parameter space and is stabilized principally at the expense of the 2D and 3D spiral orders, and lesser so at the cost of the stripe and planar AF orders. The phase boundaries and the pitch vectors of the helimagnetic orders are found to be strongly renormalized compared to their classical values, however, no new magnetic orders are found to be stabilized by quantum fluctuations. We estimate the critical magnetic ordering temperature $T_{c}$ of the N\'eel and FM orders in the $J_{1}$\textendash$J_{2}$ Heisenberg model and compare our findings against Quantum Monte Carlo estimates for nonfrustrated case of nearest-neighbor FM and AF couplings only, and previously obtained high-temperature series expansion estimates in the frustrated regime.  

The paper is organized as follows: In Sec.~{\ref{sec:methods}}, we briefly introduce the main methods which are employed in this paper. These are the Luttinger-Tisza approach (Sec.~\ref{sec:LT}) which is used to determine the classical phase diagram and the PFFRG method (Sec.~\ref{sec:frg}) which is employed to map out the quantum phase diagram for $S=1/2$. The following Sec.~\ref{sec:results} presents the results of this study, wherein Sec.~\ref{sec:classical} (Sec.~\ref{sec:quantum}) discuss the classical (quantum) phase diagrams. Sec.~\ref{sec:tc} is devoted to an analysis of critical magnetic ordering temperatures. Finally, we summarize our findings and present an outlook for future studies in Sec.~\ref{sec:conclusion}. 
In Appendix \ref{app:ccm} we provide a brief illustration of the coupled-cluster method (CCM) that we use to complement our calculations. In Appendix \ref{app:qmc} we report some details on the quantum Monte Carlo simulations of the finite-temperature behavior in the unfrustrated regime.

\section{Methods}\label{sec:methods}

\subsection{Luttinger-Tisza method}\label{sec:LT}
The classical limit of a quantum spin model can be obtained by replacing all spin operators on the lattice by unit vectors. For general Heisenberg-type interactions on the BCC lattice, the classical spin Hamiltonian is thus
\begin{equation}\label{eqn:Hamclass}
 {\cal H} = \sum_{i,j} J(|\mathbf{R}_{i}-\mathbf{R}_{j}|) \mathbf{S}_{i} \cdot \mathbf{S}_{j},
\end{equation}
where the Heisenberg spin operators $\mathbf{\hat{S}}_{i}$, such as in Eq.~\eqref{eqn:heis-ham}, are reduced to standard three-component unit vectors $\mathbf{S}_{i}$. Here, $\mathbf{R}_{i}$ is the position of the lattice site $i$. Formally, this limit can be understood as normalization of the spin operators by the total angular momentum $\sqrt{S(S+1)}$, and subsequently taking the limit $S \to \infty$~\cite{Millard-1971,Lieb-1973}. The Luttinger-Tisza method~\cite{Luttinger-1946, Luttinger-1951, Kaplan-2007} is an approach to find the approximate groundstate of the classical model in Eq.~\eqref{eqn:Hamclass} by replacing the unit length constraint for each classical spin vector with a global constraint $\sum_{i}|\mathbf{S}_{i}^{2}|=S^{2}N$, called a \emph{weak constraint}. Here, $N$ denotes the total number of spins in the system. This approximation, in principle, allows for local fluctuations in the spin length as only the average of the local moments is fixed. On Bravais lattices, such as the BCC lattice, however, the ground state subject to the weak constraint automatically fixes $|\mathbf{S}_{i}^{2}|=S^{2}\ \forall~i$, which renders the Luttinger-Tisza method exact on Bravais lattices.

To understand this and also solve the weakly constrained problem we switch to reciprocal space where the classical Heisenberg Hamiltonian Eq.~\eqref{eqn:Hamclass} reads
\begin{equation}
 {\cal H} = \sum_{\mathbf{k}} J(\mathbf{k}) \mathbf{S}(\mathbf{k}) \cdot \mathbf{S}(-\mathbf{k}).
\end{equation}
Here, we have used the Fourier transform of the spin configuration 
\begin{equation}
 \mathbf{S}(\mathbf{k}) = \frac{1}{\sqrt{N}} \sum_i e^{-\imath \mathbf{k} \cdot \mathbf{R}_i}  \mathbf{S}_{i}.
\end{equation}
An analogous expression also gives the Fourier transform of the interaction $J(\mathbf{k})$.

On a Bravais lattice, the normalized Fourier modes in real space are planar spin spirals given by 
\begin{equation}
 \mathbf{S}_{i} = R (\pm\cos(\mathbf{k} \cdot \mathbf{R}_i), \sin(\mathbf{k} \cdot \mathbf{R}_i), 0),
\end{equation}
where $R \in O(3)$ is an arbitrary rotation and reflection matrix allowed by the symmetry of the Heisenberg model. The choice of sign in the first component of $\mathbf{S}_{i}$ reflects the chirality of the spiral and is not fixed within the Heisenberg model alone and instead requires the presence of terms anisotropic in spin space such as dipolar interactions. The classical ground state is obtained by minimizing $J(\mathbf{k})$ with respect to $\mathbf{k}$ within the first Brillouin zone, and the real space spin configuration is then given by the corresponding spin spiral. For the case of the $J_1$-$J_2$-$J_3$ Heisenberg model on the BCC lattice in Eq.~(\ref{eqn:heis-ham}) we obtain
\begin{widetext}
\begin{align}
J(\mathbf{k}) = 8 J_1\cos\bigg{(}\frac{k_x}{2}\bigg{)} \cos\bigg{(}\frac{k_y}{2}\bigg{)} \cos\bigg{(}\frac{k_z}{2}\bigg{)}
        + &2 J_2 (\cos(k_x)+\cos(k_y)+\cos(k_z)) \nonumber \\
        + &4 J_3 (\cos(k_x)\cos(k_y)+\cos(k_y)\cos(k_z)+\cos(k_x)\cos(k_z))\label{eq:fourierj}
\end{align}
\end{widetext}
where, $k_x$, $k_y$ and $k_z$ are the three components of the wavevector $\mathbf{k}$. It is possible to analytically carry out the minimization of $J(\mathbf{k})$, and the wave vector corresponding to the minima in $\mathbf{k}$-space is termed as the \emph{ordering} wavevector, and subsequently denoted by $\mathbf{q}$. The ordering wave vectors can be unique \emph{or} degenerate for a particular ordered state, but are always distinct for two different magnetic orders. Hence, an ordered state can be uniquely specified by its $\mathbf{q}$ vector(s). 

Employing this scheme, we obtain \emph{all} the different classical ground states stabilized in the $J_1$\textendash$J_2$\textendash$J_3$ BCC Heisenberg model. The analytical minimizations are performed only along the high symmetry lines of the first Brillouin zone of the BCC lattice, where all the $\mathbf{q}$'s for the different ground states are found to be located. Away from these lines numerical minimizations are done only for completeness. Mapping out the respective ground states in the $J_1$\textendash$J_2$\textendash$J_3$ parameter space permits us to build the complete analytical phase diagram of the Heisenberg model on the BCC lattice. We also obtain the analytical expressions for the phase boundaries between different magnetically ordered ground states, and the order of the phase transitions.

\subsection{Pseudofermion functional renormalization group method}\label{sec:frg}

We now briefly introduce the PFFRG method which is used to calculate the quantum ($S=1/2$) phase diagram of the system. The first key step of this approach~\cite{Reuther-2010,Reuther2011,Reuther2011_1,Reuther2011_2,Reuther2011_3,Thomale-2014} is to re-express the spin Hamiltonian, e.g., Eq.~\eqref{eqn:heis-ham}, in terms of Abrikosov pseudofermions using $\mathbf{\hat{S}}_{i}=\frac{1}{2}\sum_{\alpha,\beta}\hat{c}^{\dagger}_{i,\alpha}\pmb{\sigma}_{\alpha\beta}\hat{c}_{i,\beta}$~\cite{Abrikosov-1965}, where $\alpha$, $\beta=\uparrow$ or $\downarrow$, and $\hat{c}^{\dagger}_{i,\alpha}$ ($\hat{c}_{i,\alpha}$) are the pseudofermion creation (annhilation) operators, and $\pmb{\sigma}$ is the Pauli matrix vector. The introduction of pseudofermions leads to an artificial enlargement of the Hilbert space, which, apart from the local physical spin states \textbar$\uparrow\rangle=|1,0\rangle$ and \textbar$\downarrow\rangle=|0,1\rangle$ also contains empty and doubly occupied states, $|0,0\rangle$ and $|1,1\rangle$, respectively, carrying zero spin (here, the notation $|n_\uparrow,n_\downarrow\rangle$ indicates the occupations of the $\uparrow$ and $\downarrow$ modes). The problem of possible spurious contributions from unphysical $S=0$ states in the PFFRG can be cured by adding level repulsion terms $-A\sum_i {\bf S}_i^2$ to the Hamiltonian which, upon choosing $A$ sufficiently large (and positive), energetically separate the Hilbert spaces of the \textbar$\uparrow\rangle$, \textbar$\downarrow\rangle$ and $|0,0\rangle$, $|1,1\rangle$ states, respectively~\cite{Baez2017}. For most practical purposes (such as an application to the models considered here), it is sufficient to set $A=0$ since the single occupancy constraint is already naturally fulfilled in the ground state~\cite{Baez2017}. This is because an unphysical occupation may be viewed as a vacancy in the spin lattice associated with a finite excitation energy.

\begin{figure}
\includegraphics[width=0.95\columnwidth]{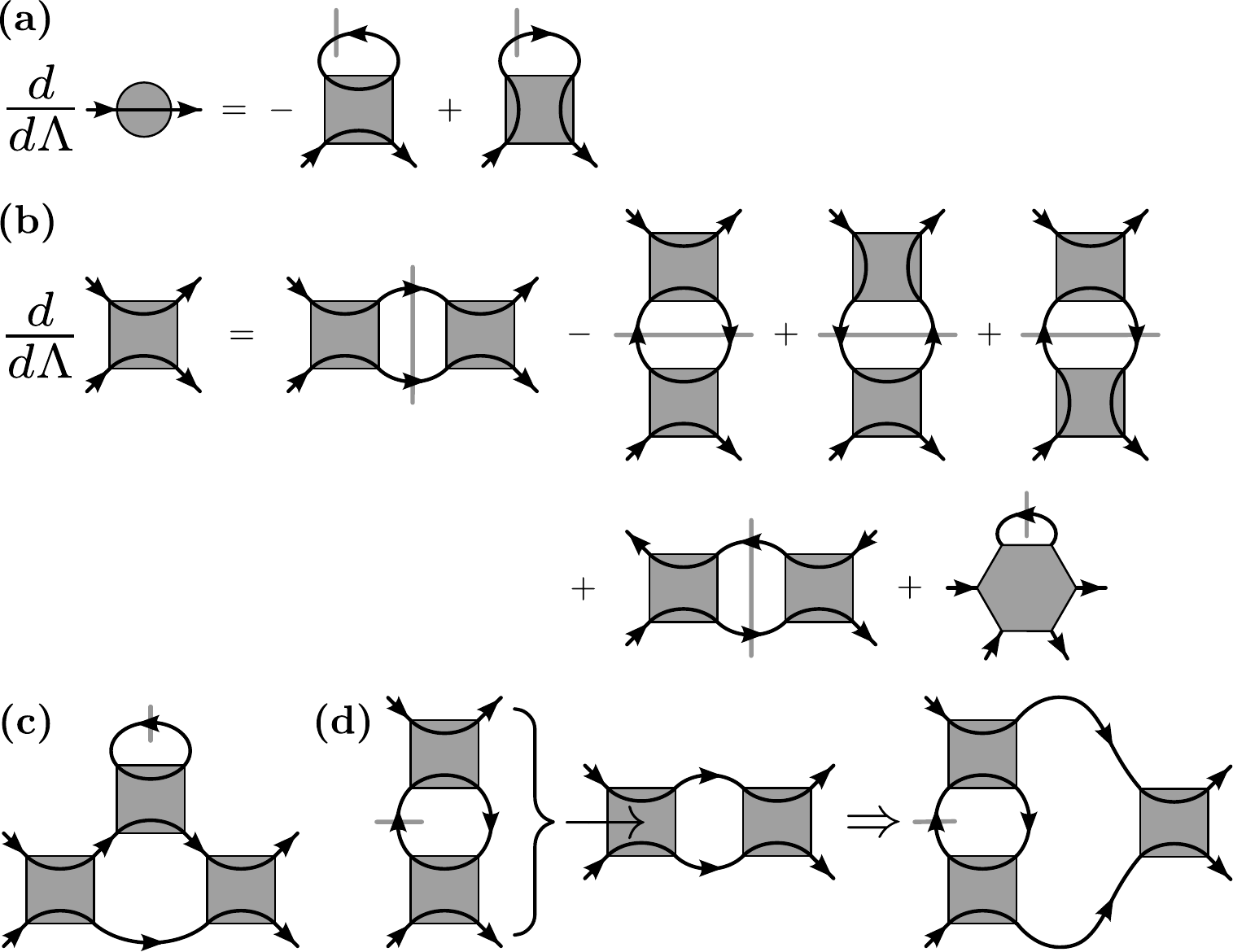}
\caption{Diagrammatic representation of the FRG equations for (a) the self energy $\Sigma$ and (b) the two-particle vertex $\Gamma$. Fermionic propagators are drawn as lines with an arrow, where gray slashes indicate that a $\Lambda$ derivative acts on the cutoff function (resulting in so-called single-scale propagators). Gray slashes crossing two propagators indicate that the derivative acts on the product of both propagators. (c) Within the Katanin truncation the contracted three-particle vertex [last term on the right hand side in (b)] is approximated by the depicted diagram representing a self-energy correction to the two-particle vertex flow. Note that the self-energy correction is only depicted for the particle-particle channel while further diagrammatic contributions can be constructed from the other channels. (d) Different interaction channels may be inserted into each other to yield additional two-loop diagrammatic corrections. Note that only one particular nested two-loop diagram is shown corresponding to one possible choice of inserting two interaction channels into each other.}\label{fig:diagrams}
\end{figure}

The resulting fermionic theory is then equipped with a step-like infrared frequency cutoff $\Lambda$ suppressing the bare fermion propagator between $\omega=-\Lambda$ and $\omega=\Lambda$ on the Matsubara axis. This manipulation generates a $\Lambda$-dependence of all $m$-particle vertex functions which may be formulated as an exact but infinite hierarchy of coupled differential equations~\cite{Metzner2012,Platt2013}, where the ones for the self energy $\Sigma$ and for the two-particle vertex, $\Gamma_2$, are illustrated in Fig.~\ref{fig:diagrams}(a) and Fig.~\ref{fig:diagrams}(b). To be amenable to numerical solutions, this hierarchy of equations needs to be truncated, which, for the results presented below, amounts to approximating the three-particle vertex, $\Gamma_3$, using two different schemes: \\
\indent(i) \emph{One-loop plus Katanin scheme:} This approach has been widely used and has proven suitable to capture the right balance between ordering tendencies and quantum fluctuations. It will be applied in most calculations presented below. Within this scheme, the contracted three-particle vertex in Fig.~\ref{fig:diagrams}(b) takes the form of Fig.~\ref{fig:diagrams}(c) which corresponds to a self-energy insertion in the interaction channels of the two-particle vertex flow. The crucial benefit of this approximation as compared to neglecting $\Gamma_3$ completely is that it guarantees the full feedback of the self-energy into the two-particle vertex flow, hence, leading to a fully self-consistent RG scheme. From a different perspective, it can be shown that the one-loop plus Katanin scheme exactly sums up all diagrammatic contributions separately in the large $S$ limit~\cite{Baez2017} and in the large $N$ limit (where in the latter case the spins' symmetry group is promoted from SU(2) to SU($N$)) \cite{Buessen2018}. This ensures that magnetically ordered phases (as typically encountered for $S\rightarrow\infty$) as well as disordered phases (obtained for $N\rightarrow\infty$) may both be faithfully described. The three-particle terms which are neglected within the one-loop plus Katanin scheme can be shown to be subleading in both $1/S$ and $1/N$.\\
\indent(ii) \emph{Two-loop scheme:} This approach adds further corrections to the three-particle term such as those shown in Fig.~\ref{fig:diagrams}(d). In this diagram, different two-particle interaction channels are inserted into each other resulting in effective two-loop contributions. It should also be noted that in similarity to the Katanin scheme in (i), self-consistency again requires the full feedback of self-energy into such nested diagrams which even generates certain three-loop contributions (see \cite{Rueck2018} for details). All these corrections ensure that the aforementioned subleading terms in $1/S$ and $1/N$ are better approximated which allows for a more accurate investigation of quantum critical parameter regions where the detailed interplay between magnetic ordering and quantum fluctuations becomes crucial. While this may generally lead to shifted phase boundaries compared to the scheme in (i) it has been shown in Ref.~\cite{Rueck2018} that such shifts turn out to be rather small. Another benefit of the two-loop scheme is that it determines N\'eel/Curie temperatures more accurately \cite{Rueck2018} which is also the context in which it will be applied below.

Up to fermionic contractions, the two-particle vertex [either calculated via (i) or (ii)] is the diagrammatic representation of the static (i.e., imaginary time-integrated) spin correlator given by
\begin{equation}
C_{ij}=C_{ij}^{zz}=\int_0^\infty d\tau\langle\hat{S}_i^z(\tau)\hat{S}_j^z(0)\rangle,\label{correlator}
\end{equation}
where $\hat{S}_i^\mu(\tau)=e^{\tau\hat{\mathcal{H}}}\hat{S}_i^\mu e^{-\tau\hat{\mathcal{H}}}$. Since the Heisenberg model is spin-rotation invariant all diagonal components $C_{ij}^{\mu\mu}$ of the spin correlator are identical. Without loss of generality we have chosen the $zz$-component here. Within PFFRG, the thermodynamic limit is approached by calculating the correlators $C_{ij}$ only up to a maximal distance between sites $i$ and $j$. Fourier-transforming $C_{ij}$ into momentum space, we then obtain the static susceptibility $\chi^\Lambda({\bf k})$ as a function of $\Lambda$ which represents the central physical outcome of this approach. While the Fourier-transform generally allows to access a continuous set of wave vectors {\bf k} within the Brillouin zone, the restriction to a finite set of correlators $C_{ij}$ limits the number of harmonics in the Fourier-sums and, therefore, smoothens sudden changes in the susceptibility. In the present study of the $J_1$\textendash$J_2$\textendash$J_3$ model on the BCC lattice, we have set the maximal length of spin correlators to be equal to 10 nearest-neighbor lattice spacings, which incorporates a total of $2331$ correlated sites, producing well converged results with a proper {\bf k}-space resolution. We, furthermore, approximate the frequency dependences of the vertex functions by discrete grids containing $64$ or $100$ points for each frequency variable. The number of coupled differential equations for the above given system size and 64 frequencies is (i) \num{611057728}\textemdash without using any point group symmetry and (ii) \num{1537568}\textemdash upon exploiting the complete $O_{h}$ point group~\cite{LandauQM} symmetries, while for a calculation with $100$ frequencies, we have (i) \num{2331000100}\textemdash without symmetries and (ii) \num{5801300}\textemdash with symmetries. When a system spontaneously develops magnetic order, the susceptibility $\chi^{\Lambda}({\bf k})$ shows a sharp increase at the corresponding wave vector {\bf q} upon decreasing $\Lambda$ and eventually the flow becomes unstable. The $\Lambda$ values at which this breakdown takes place can be associated with the critical ordering temperature via the relation $T_c=(\frac{\pi}{2})\Lambda_c$ for $S=1/2$~\cite{Iqbal-2016a,Iqbal-2019a}. In contrast, a smooth flow of the susceptibility down to $\Lambda\rightarrow0$ indicates a magnetically disordered state. After the initial applications of the PFFRG in two dimensions, it has subsequently been applied with much success to three-dimensional systems~\cite{Iqbal-2016a,Balz-16,Iqbal-2019a,Buessen-16a,Iqbal-2017,Chillal-2017,Iqbal-2018,Buessen-2018}. For further details about the PFFRG procedure, its subsequent refinements, and expansions to handle a larger class of magnetic Hamiltonians we refer the reader to Refs.~\cite{Reuther-2010,Reuther2011,Reuther2011_2,Iqbal-2016a,Poilblanc-2016,Hering-2017,Keles-2018,Buessen-2017,Roscher-2017,Iqbal-2015}.

\section{Results}\label{sec:results}

\subsection{Classical Phase diagram}\label{sec:classical}

\begin{figure*}
\includegraphics[width=1\columnwidth]{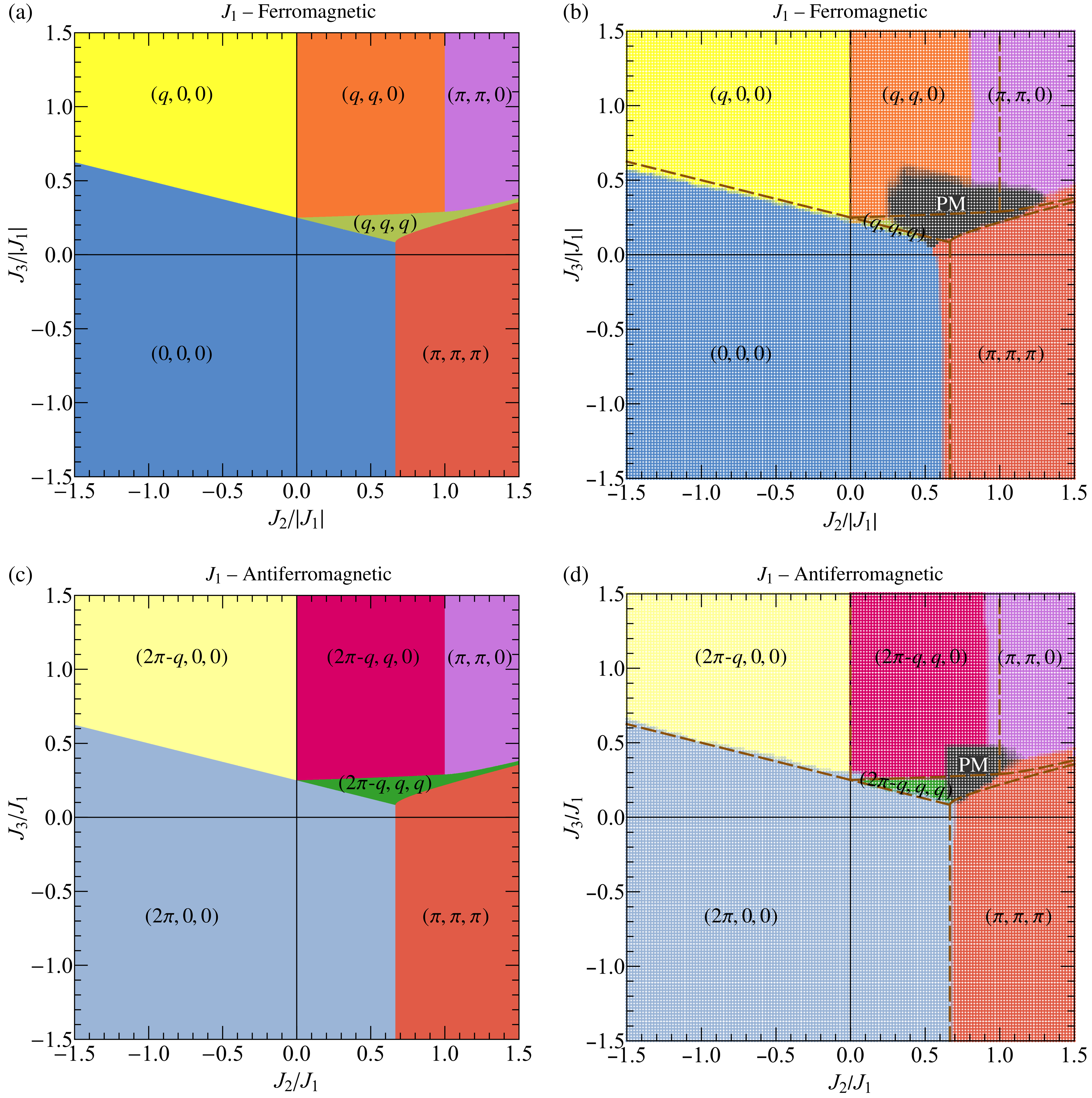}
\caption{The classical phase diagrams of the $J_{1}$\textendash$J_{2}$\textendash$J_{3}$ Heisenberg model on the BCC lattice for (a) ferromagnetic $J_{1}$ and (c) antiferromagnetic $J_{1}$. The corresponding $S=1/2$ quantum phase diagrams are shown in (b) for ferromagnetic $J_{1}$ and (d) for antiferromagnetic $J_{1}$. Note the change in the phase boundaries and the appearance of a paramagnetic phase in both quantum phase diagrams. The (brown) dashed lines overlaid in (b) and (d) denote the classical phase boundaries, and are meant as an aid to visualize the modifications in the quantum phase diagram with respect to the corresponding classical one.}\label{fig:cpd-qpd}
\end{figure*}

\floatsetup[table]{capposition=bottom}
\begin{table*}
\setlength{\tabcolsep}{6pt}
\setlength\extrarowheight{2pt}
\centering
\begin{tabular}{m{2.3cm}m{5cm}m{2.3cm}m{6cm}}
\hline\hline
Pitch vector ($\mathbf{q}$) & Component $q$ & Degeneracy of $\mathbf{q}$ in the first Brillouin zone & Energy $\frac{E}{N S^2}$\\
\hline
$(q,0,0)$ & \multirow{2}{*}{$2\cos^{-1}\Big{(}\frac{|J_1|}{J_2+4 J_3}\Big{)}$} &\multirow{2}{*}{6\textendash fold} & \multirow{2}{*}{$J_2-2 J_3 - \frac{2 (J_1)^2}{J_2+4 J_3}$}\\
$(2\pi-q,0,0)$ &  \\  \hline
$(q,q,0)$ & \multirow{2}{*}{$\cos^{-1}\Big{(}\frac{|J_1|-J_2-2 J_3}{2 J_3}\Big{)}$}& 12\textendash fold & \multirow{2}{*}{$-J_2 - 2 J_3- \frac{(-|J_1|+J_2)^2}{2 J_3}$}\\
$(2\pi-q,q,0)$ & & 24\textendash fold  \\ \hline
$(q,q,q)$ & \multirow{2}{*}{$2 \cos^{-1}\Big{(}\frac{|J_1|+\sqrt{(J_1)^2-32 J_3 (J_2-4 J_3)}}{16 J_3}\Big{)}$}&8\textendash fold & $-\frac{(J_1)^2 (u +16  J_3 (4 J_3 + J_2))+ |J_1| u^{\frac{3}{2}} +384 (J_2)^2 (J_3)^2}{1024 (J_3)^3}$\\
$(2\pi-q,q,q)$ & & 24\textendash fold & $u=(J_1)^2 + 32 J_3 (4J_3-J_2)$\\ \hline
$(0,0,0)$ & &\multirow{2}{*}{1\textendash fold} & \multirow{2}{*}{$-4 |J_1| +3 J_2 + 6 J_3$}\\
$(2\pi,0,0)$ & & \\  \hline
$(\pi,\pi,0)$ & & 6\textendash fold & $- J_{2}-2 J_{3}$\\ \hline
$(\pi,\pi,\pi)$ & & 1\textendash fold & $-3 J_{2} + 6 J_{3}$\\
\hline\hline
\end{tabular}
\caption{The pitch vectors of the helimagnetic orders in the classical $J_1$\textendash$J_2$\textendash$J_3$ Heisenberg model on the BCC lattice together with their degeneracy in the first Brillouin zone and the energy per spin in the corresponding ground state.}
\label{tab:Ordering-Vec-Classical}
\end{table*}

\begin{table*}
\setlength{\tabcolsep}{5pt}
\setlength\extrarowheight{2pt}
\begin{tabular}{m{2.3cm}m{2.3cm}m{8.5cm}m{2cm}}
\hline\hline
Phase I & Phase II & Equation for the phase boundary & Type of Phase Transition\\
\hline
$(0,0,0)$ & $(q,0,0)$ & \multirow{2}{*}{$J_{2}=|J_{1}|-4 J_{3}$} & \multirow{2}{*}{2nd Order}\\
$(2\pi,0,0)$ & $(2\pi-q,0,0)$ &  & \\
\hline
$(0,0,0)$ & $(q,q,q)$ & \multirow{2}{*}{$J_{2}=|J_{1}|-4 J_{3}$} & \multirow{2}{*}{2nd Order}\\
$(2\pi,0,0)$ & $(2\pi-q,q,q)$ &  & \\
\hline
$(0,0,0)$ & \multirow{2}{*}{$(\pi,\pi,\pi)$} & \multirow{2}{*}{$J_{2}=\frac{2}{3}|J_{1}|$} & \multirow{2}{*}{1st Order}\\
$(2\pi,0,0)$ &  & &  \\
\hline
$(q,0,0)$ & $(q,q,0)$ & \multirow{2}{*}{$J_{2}=0$} & \multirow{2}{*}{1st Order}\\
$(2\pi-q,0,0)$ & $(2\pi-q,q,0)$ &  &\\
\hline
$(q,q,0)$ & \multirow{2}{*}{$(\pi,\pi,0)$} & \multirow{2}{*}{$J_{2} = |J_{1}|$}& \multirow{2}{*}{2nd Order} \\
$(2\pi-q,q,0)$ &  & & \\
\hline
$(q,q,0)$ & $(q,q,q)$ & \multirow{2}{*}{$J_{2} = \frac{56 |J_{1}|(J_{3})^{2} - 11 (J_{1})^{2}J_{3} - 32 (J_{3})^{3} - \sqrt{|J_{1}| (J_{3})^{2} (5 |J_{1}| - 16 J_{3})^3}}{8 (J_{3})^{2}}$} & \multirow{2}{*}{1st Order}\\
$(2\pi-q,q,0)$ & $(2\pi-q,q,q)$ & & \\
\hline
$(q,q,q)$ & \multirow{2}{*}{$(\pi,\pi,0)$} & \multirow{2}{*}{$J_{2} = \frac{(J_{1})^{4}+4416 (J_{1})^{2} (J_{3})^{2}+((J_{1})^{2}+48 (J_{3})^{2}) \sqrt[3]{u}+\left(u\right)^{2/3}+36864 \left(J_{3}\right)^{4} }{108 {J_3} \sqrt[3]{u}}$} & \multirow{2}{*}{1st Order}\\
$(2\pi-q,q,q)$ &  & & \\
\hline
$(q,q,q)$ & \multirow{2}{*}{$(\pi,\pi,\pi)$} & \multirow{2}{*}{$J_{2} = \frac{\left(J_{1}\right)^{2}+144 \left(J_{3}\right)^{2}}{36 {J_{3}}}$} & \multirow{2}{*}{1st Order}\\
$(2\pi-q,q,q)$ &  & &  \\
\hline\hline\\
\multicolumn{4}{l}{with $u = (J_{1})^{6}-10872 (J_{1})^{4} (J_{3})^{2} - 2709504 (J_{1})^{2} (J_{3})^{4}+108 \sqrt{3} J_{3} \sqrt{-(J_{1})^{2} \left(12 (J_{3})^{2}-(J_{1})^{2}\right)\left(512 (J_{3})^{2}-(J_{1})^{2}\right)^{3}}+7077888 (J_{3})^{6}$}
\end{tabular}
\caption{For the classical phase diagram of the $J_{1}$\textendash$J_{2}$\textendash$J_{3}$ Heisenberg model on the BCC lattice, we provide the analytical expression for the different phase boundaries between the members of column Phase I and Phase II, and the order of the corresponding phase transitions.}
\label{tab:Boundaries-Classical}
\end{table*}

We begin by presenting the classical ground state phase diagram, i.e., at $T=0$ and a detailed analysis of the magnetic orders in the classical $J_{1}$\textendash$J_{2}$\textendash$J_{3}$ Heisenberg model on a BCC lattice as obtained from the Luttinger-Tisza method [Sec.~\ref{sec:LT}].

\subsubsection{Ferromagnetic $J_1$}
The classical phase diagram in the $J_{1}$\textendash$J_{2}$\textendash$J_{3}$ parameter space with FM $J_{1}$ is shown in Fig.~\ref{fig:cpd-qpd}(a). It is host to six different types of magnetic orders; three incommensurate coplanar spiral structures and three collinear orders. Starting with both FM $J_{1}$ and $J_{2}$ we trivially find a FM ground state [Fig.~\ref{fig:mag-structure}(a)]. The inclusion of an AF $J_{3}$ coupling above a critical value $J_{3}^{c}=\frac{1}{4}(|J_{1}|+|J_{2}|)$ destabilizes the FM state~\cite{Kaplan-1959}, via a 2nd order phase transition, into an incommensurate 1D spiral [Fig.~\ref{fig:mag-structure}(e)] with a pitch vector $\mathbf{q}=(q,0,0)$ with $q$ given in 
Table~\ref{tab:Ordering-Vec-Classical}. This pitch vector is 6\textendash fold degenerate within the first Brillouin zone. It is important to emphasize that the spiral state is only governed by \emph{one} of these symmetry equivalent pitch vectors, i.e., superpositions are not possible as they would violate the classical length constraint, and hence the degeneracy remains discrete. This 1D spiral structure is stabilized purely by a FM $J_2$ interaction. Indeed, along the line $J_{2}=0$, there is a 1st order phase transition to a 2D incommensurate spiral [Fig.~\ref{fig:mag-structure}(f)] with a pitch vector $\mathbf{q}=(q,q,0)$ with $q$ given in 
Table~\ref{tab:Ordering-Vec-Classical}. Similar to the 1D spiral, the ground state in this phase is determined only by \emph{one} of the $12$ symmetry equivalent $(q,q,0)$-type pitch vectors. Upon increasing $J_{2}$, we observe that the value of $q$ continuously evolves towards $\pi$. At the line $J_{2}/|J_{1}|=1$ and above a critical $J_{3}^{c}/|J_{1}|\approx0.29$ 
[see Table~\ref{tab:Boundaries-Classical} for an analytical expression of the phase boundaries], there is a 2nd order phase transition to a planar AF order [Fig.~\ref{fig:mag-structure}(d)] with $\mathbf{q}=(\pi,\pi,0)$~\cite{Utsumi-1977}. In contrast to the incommensurate orders discussed above, the pitch vector of the planar AF is half of a reciprocal lattice vector, i.e., $2\mathbf{q}\equiv\mathbf{0}$. As pointed out by Villain~\cite{Villain-1977}, this characteristic allows the ground state to be composed of all six symmetry equivalent pitch vectors, namely, $(\pi,\pi,0)$, $(\pi,0,\pi)$, $(0,\pi,\pi)$, $(\pi,-\pi,0)$, $(-\pi,0,\pi)$, and $(0,-\pi,\pi)$. All six pitch vectors satisfy the property $\sin(\mathbf{q}\cdot\mathbf{R}_{i})=0$ at every lattice site. Therefore, the general ground state can be written as

\begin{widetext}
\begin{gather}
\begin{split}
\mathbf{S}_{i} = S\{&\mathbf{a}\cos[(\pi,\pi,0)\cdot\mathbf{R}_{i}]+\mathbf{b}\cos[(\pi,0,\pi)\cdot\mathbf{R}_{i}]+\mathbf{c}\cos[(0,\pi,\pi)\cdot\mathbf{R}_{i}] \\ 
&+\mathbf{d}\cos[(\pi,-\pi,0)\cdot\mathbf{R}_{i}]+\mathbf{e}\cos[(-\pi,0,\pi)\cdot\mathbf{R}_{i}]+\mathbf{f}\cos[(0,-\pi,\pi)\cdot\mathbf{R}_{i}]\}
\end{split}\\ \nonumber \\
\begin{split}
\mathbf{a}^{2}+\mathbf{b}^{2}+\mathbf{c}^{2}+\mathbf{d}^{2}+\mathbf{e}^{2}+\mathbf{f}^{2}&=1\\
\mathbf{a}\cdot\mathbf{b}+\mathbf{d}\cdot\mathbf{e}=\mathbf{a}\cdot\mathbf{c}+\mathbf{d}\cdot\mathbf{f}=\mathbf{a}\cdot\mathbf{e}+\mathbf{b}\cdot\mathbf{d}=\mathbf{a}\cdot\mathbf{f}+\mathbf{c}\cdot\mathbf{d}=\mathbf{b}\cdot\mathbf{c}+\mathbf{e}\cdot\mathbf{f}&=\mathbf{c}\cdot\mathbf{e}+\mathbf{b}\cdot\mathbf{f}=\mathbf{b}\cdot\mathbf{e}+\mathbf{a}\cdot\mathbf{d}+\mathbf{c}\cdot\mathbf{f}=0
\end{split}
\label{eq:pipi0}
\end{gather}
\end{widetext}

\noindent where $\mathbf{a}$, $\mathbf{b}$, $\mathbf{c}$, $\mathbf{d}$, $\mathbf{e}$, $\mathbf{f}$ are arbitrary vectors constrained by Eq.~\eqref{eq:pipi0} which normalizes the spin length at each site. The arbitrary vectors defining the spin configuration have in total $18$ continuous degrees of freedom, and are subject to $8$ constraints. Accounting for the global spin rotation invariance (2 degrees of freedom) of the Heisenberg model, the \emph{continuous} ground state manifold of the planar AF order is $8$-dimensional. This is in contrast to the other magnetic orders in the $J_{1}$\textendash$J_{2}$\textendash$J_{3}$ parameter space which feature only a $n$-fold \emph{discrete} degeneracy 
[see Table~\ref{tab:Ordering-Vec-Classical}]. A similar enhancement for the available degrees of freedom is also found for the stripe AF phase on the square lattice~\cite{Danu-2016} and other cubic lattice systems~\cite{Ignatenko-2016}.

Upon decreasing the value of $J_{3}$, we find that the interplay between AF $J_{2}$ and $J_{3}$ couplings, leads to the appearance of an incommensurate 3D spiral [Fig.~\ref{fig:mag-structure}(g)] in a sliver of parameter space. This state also continuously evolves from the FM state via a 2nd order phase transition, however, its transition into the $(q,q,0)$ and $(\pi,\pi,0)$ states is of 1st order. Its pitch vector $\mathbf{q}=(q,q,q)$ is 8\textendash fold degenerate, but again, only \emph{one} of them is present in any given ground state. Lowering the AF $J_{3}$ coupling even further, and for $J_{2}/|J_{1}|>2/3$, the collinear stripe order [Fig.~\ref{fig:mag-structure}(c)] with a wave vector $\mathbf{q}=(\pi,\pi,\pi)$ is stabilized. This state is composed of two interpenetrating simple cubic lattices which are N\'eel ordered. For any spin on a given sublattice, all its nearest-neighbor spins residing on the other sublattice add up to zero. Hence, the energy of the $(\pi,\pi,\pi)$ state is independent of the relative orientation of the two sublattices~\cite{Shender-1982}, which is thus not determined within the $J_{1}$\textendash$J_{2}$\textendash$J_{3}$ Heisenberg model. The pitch vector of this state resides at the corners of the first Brillouin zone, and is therefore unique. Since the third neighbor spins in this state are FM ordered, changing $J_{3}$ to FM only enhances its stability, and thus this state occupies the entire parameter space for $J_{2}/|J_{1}|>2/3$ and FM $J_{3}$. The phase boundary between the FM and the stripe collinear order is determined solely by the coordination number at nearest-neighbor and second-neighbor distances, and is given by $J_{2}/|J_{1}|=z_{1}/2z_{2}=2/3$.

\begin{figure*}
\includegraphics[width=1.0\columnwidth]{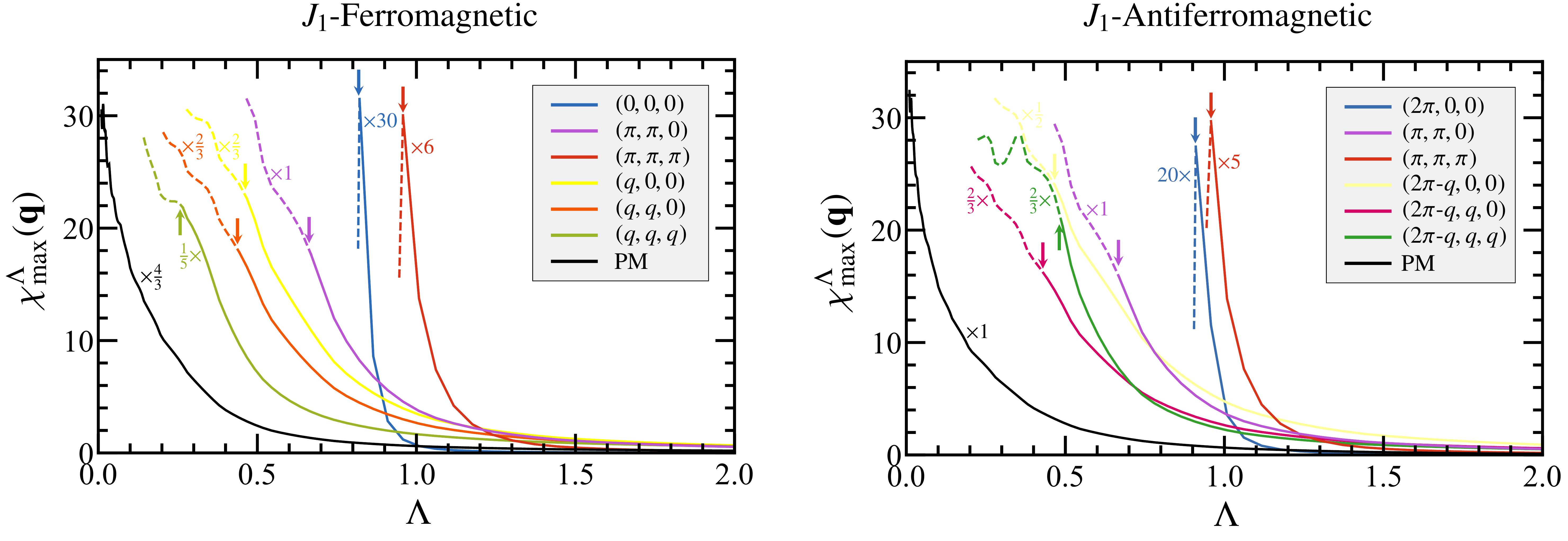}
\caption{Representative RG flows of the maximum of the magnetic susceptibilities in momentum space for the ordered regimes and the PM regimes of Fig.~\ref{fig:cpd-qpd}(b) and Fig.~\ref{fig:cpd-qpd}(d). The susceptibilities are evaluated for FM $J_{1}$ (left panel) at the following $(J_{2}/|J_{1}|,J_{3}/|J_{1}|)$ points: FM order at $(0,0)$, $(\pi,\pi,0)$ order at $(1.5,1)$, $(\pi,\pi,\pi)$ order at $(1.5,0)$, 1D spiral $\q00$ at $(-0.5,1)$, 2D spiral $\qq0$ at $(0.5,1)$, 3D spiral $\qqq$ at $(0.26,0.2)$, and PM at $(0.8,0.3)$. For AF $J_{1}$ (right panel), the magnetic susceptibilities are plotted at the same parameter points as for FM $J_{1}$. The points at which the solid lines become dashed (marked by vertical arrows) indicate an instability in the flow and express the onset of long-range magnetic order. A smooth flow down to $\Lambda\to0$ (black curves) indicates paramagnetic behavior.}\label{fig:xivslam}
\end{figure*}

\subsubsection{Antiferromagnetic $J_1$}
A change in the $J_{1}$ coupling from FM to AF is found not to alter the phase boundaries and the order of the phase transitions in the $J_{2}$\textendash$J_{3}$ parameter space, as observed in the corresponding phase diagram [Fig.~\ref{fig:cpd-qpd}(c)] and 
Table~\ref{tab:Boundaries-Classical}. This feature is most easily understood by viewing the BCC lattice as being composed of two interpenetrating simple cubic sublattices with one being positioned at the body center of the other. As $J_{2}$ and $J_{3}$ couple sites only within the sublattices, it is only the $J_{1}$ coupling which connects the two simple cubic lattices. Hence, a sign reversal of $J_1$ can be undone by flipping the spins on one of the sublattices without affecting the $J_{2}$ and $J_{3}$ couplings. Therefore, the phase boundaries remain unchanged, however, in reciprocal space this flipping amounts to a shift of the wave vector $\mathbf{q} \to (2\pi,2\pi,2\pi)-\mathbf{q}$. After folding back this wave vector into the first Brillouin zone this amounts to a shift of any \emph{one} of the wave vector components $q\to2\pi-q$. Consequently, the FM state is replaced by a N\'eel AF [Fig.~\ref{fig:mag-structure}(b)] with wavevector $\mathbf{q} = (2\pi,0,0)$~\footnote{In literature, the N\'eel AF state is also specified by the wave vector $\mathbf{q} = (2\pi,2\pi,2\pi)$. This wave vector does not reside in the first Brillouin zone, however, it is equivalent to $\mathbf{q} = (2\pi,0,0)$}. Similarly, the incommensurate 1D spiral is now characterized by the pitch vector $\mathbf{q} =(2\pi-q,0,0)$ [Fig.~\ref{fig:mag-structure}(h)] with $q$ and the degeneracy of the ground state being the same as in the FM case 
[see Table~\ref{tab:Ordering-Vec-Classical}]. Along the same lines, the pitch vector of the incommensurate 2D spiral in the AF $J_{1}$ case is given by $\mathbf{q} =(2\pi-q,q,0)$, which is now $24$\textendash fold degenerate in contrast to the $12$\textendash fold degeneracy present in the ferromagnetic $J_1$ case. Nonetheless, the ground state is still composed of only \emph{one} of these pitch vectors. This state also evolves continuously to the planar AF upon increasing $J_{2}$, and the planar AF remains unchanged compared to the FM $J_{1}$ case. The incommensurate 3D spiral now has a $24$\textendash fold degenerate pitch vector $\mathbf{q} =(2\pi-q,q,q)$. Finally, the stripe AF $\mathbf{q} =(\pi,\pi,\pi)$-phase remains unchanged.

The complete analytical expressions for the pitch vectors of the spiral orders together with their degeneracies are to be found in 
Table~\ref{tab:Ordering-Vec-Classical}. Also, in Table~\ref{tab:Boundaries-Classical} we provide the analytical expressions of all phase boundaries together with the order of the phase transitions.   

\subsection{Quantum Phase diagram}\label{sec:quantum}

\begin{figure*}
\includegraphics[width=.9\textwidth]{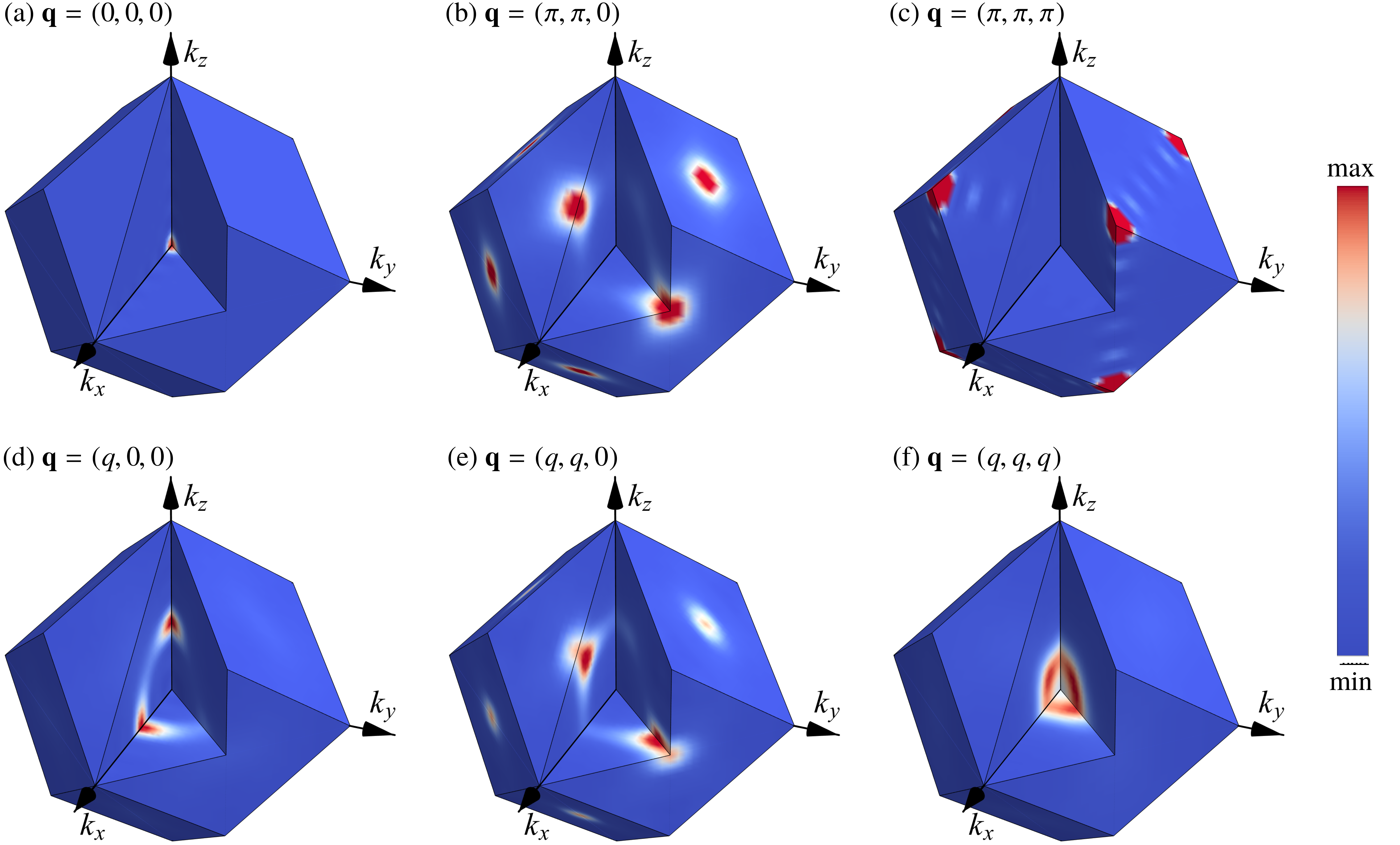}
\caption{The susceptibility profiles in the first Brillouin zone, a rhombic dodecahedron, for the $S=1/2$ $J_1$\textendash$J_2$\textendash$J_3$ Heisenberg model on a BCC lattice with FM $J_{1}$ for the magnetically ordered states. The plots are calculated for the coupling parameters quoted in the caption of Fig.~\ref{fig:xivslam}. The corresponding $\Lambda$ values are given by the respective points of instabilities as indicated by the vertical arrows in the left panel of Fig.~\ref{fig:xivslam}.}\label{fig:sus-FM}
\end{figure*}

\begin{figure*}
\includegraphics[width=.9\textwidth]{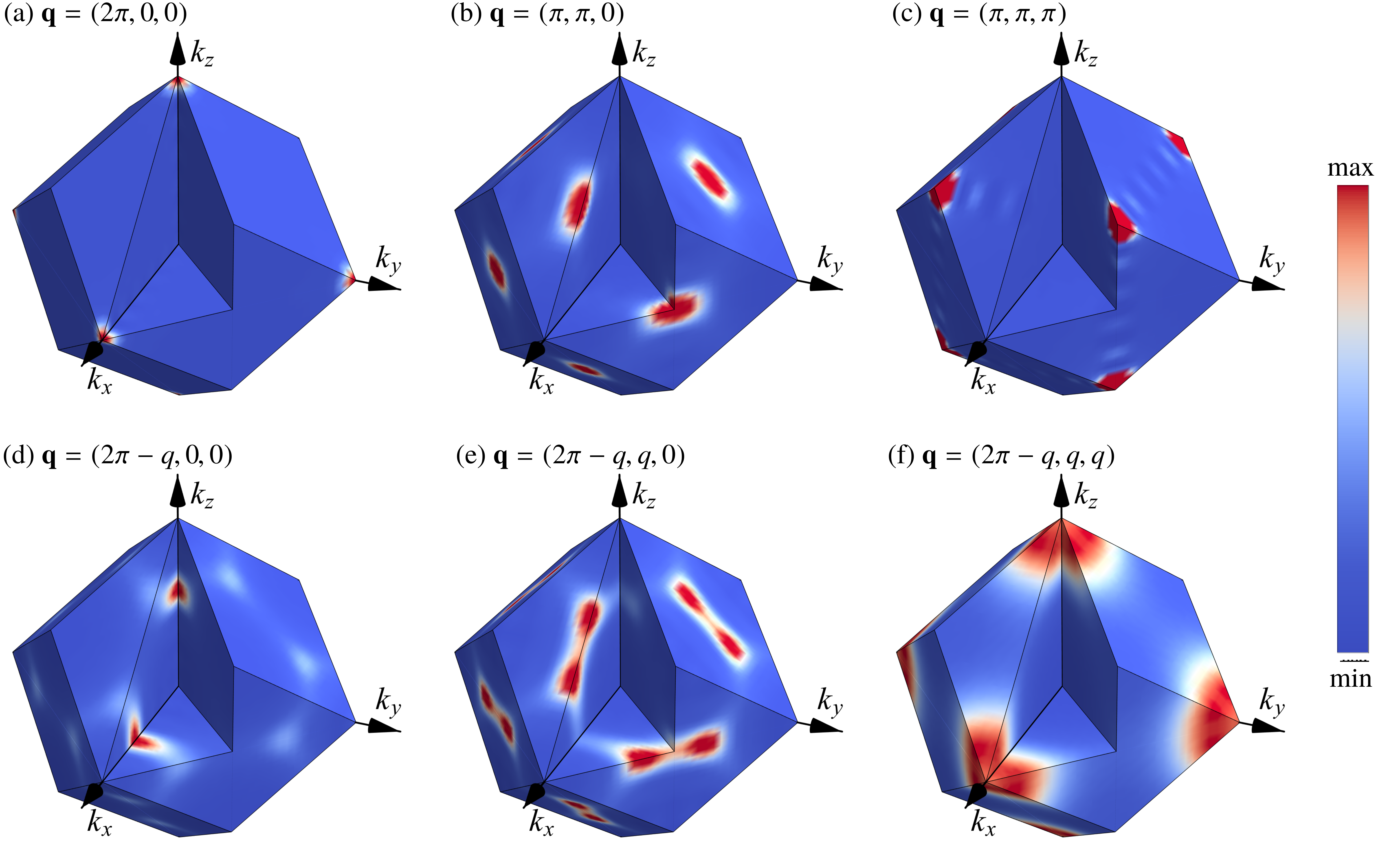}
\caption{The susceptibility profiles in the first Brillouin zone, a rhombic dodecahedron, for the $S=1/2$ $J_1$\textendash$J_2$\textendash$J_3$ Heisenberg model on a BCC lattice with AF $J_{1}$ for the magnetically ordered states. The plots are calculated for the coupling parameters quoted in the caption of Fig.~\ref{fig:xivslam}. The corresponding $\Lambda$ values are given by the respective points of instabilities as indicated by the vertical arrows in the right panel of Fig.~\ref{fig:xivslam}.}\label{fig:sus-AFM}
\end{figure*}

\begin{figure}
\includegraphics[width=0.7\columnwidth]{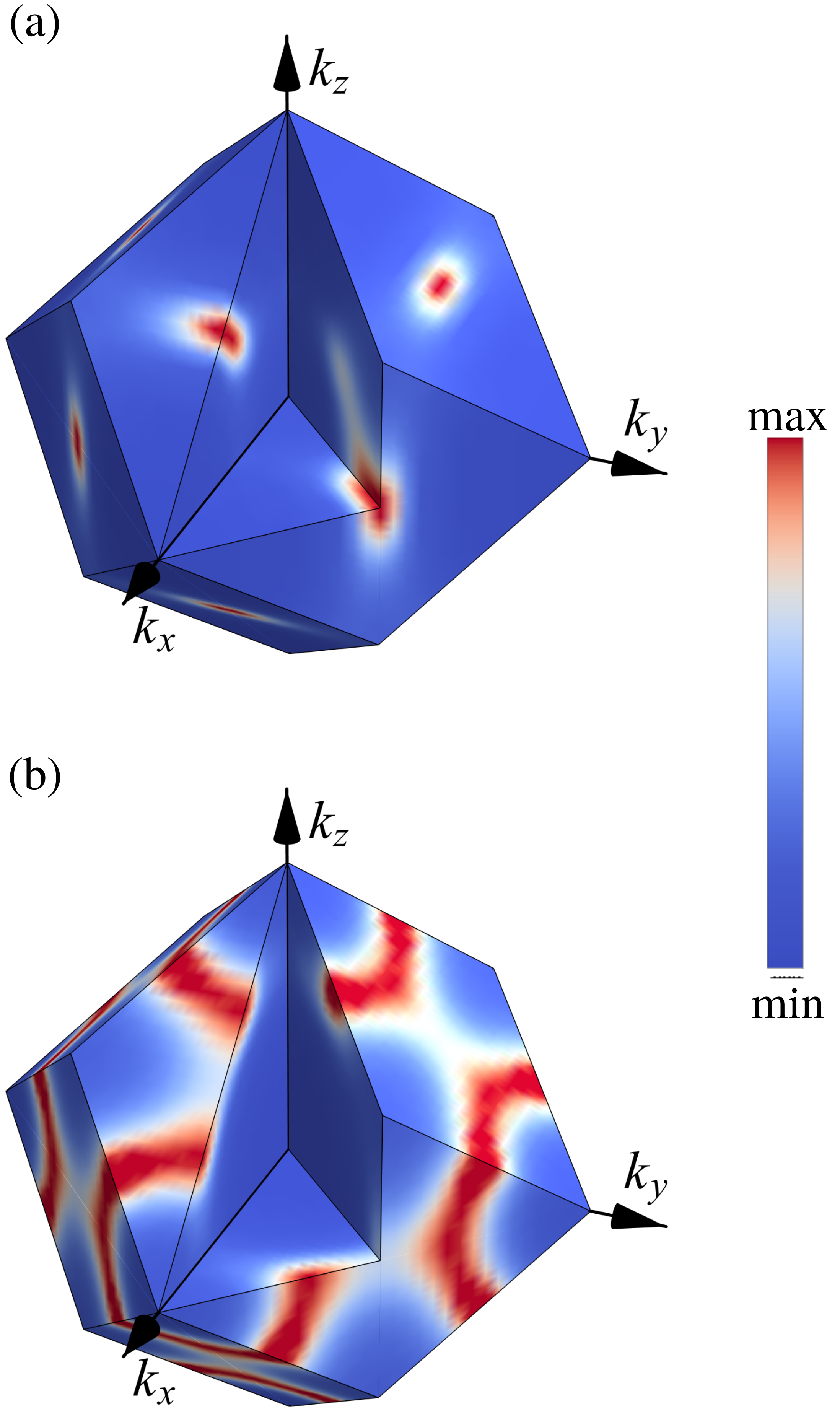}
\caption{The paramagnetic susceptibility profile evaluated at the end of the flow for (a) ferromagnetic $J_1$ and (b) antiferromagnetic $J_{1}$ for the spin-$1/2$ BCC Heisenberg model. In both plots the susceptibility profile is calculated at $(J_2/|J_{1}|,J_3/|J_{1}|)=(0.8,0.3)$. Note the softer peaks as compared to the sharp peaks in the similar plots for the ordered states.}\label{fig:sus-PM}
\end{figure}

\begin{figure}
\includegraphics[width=1.0\columnwidth]{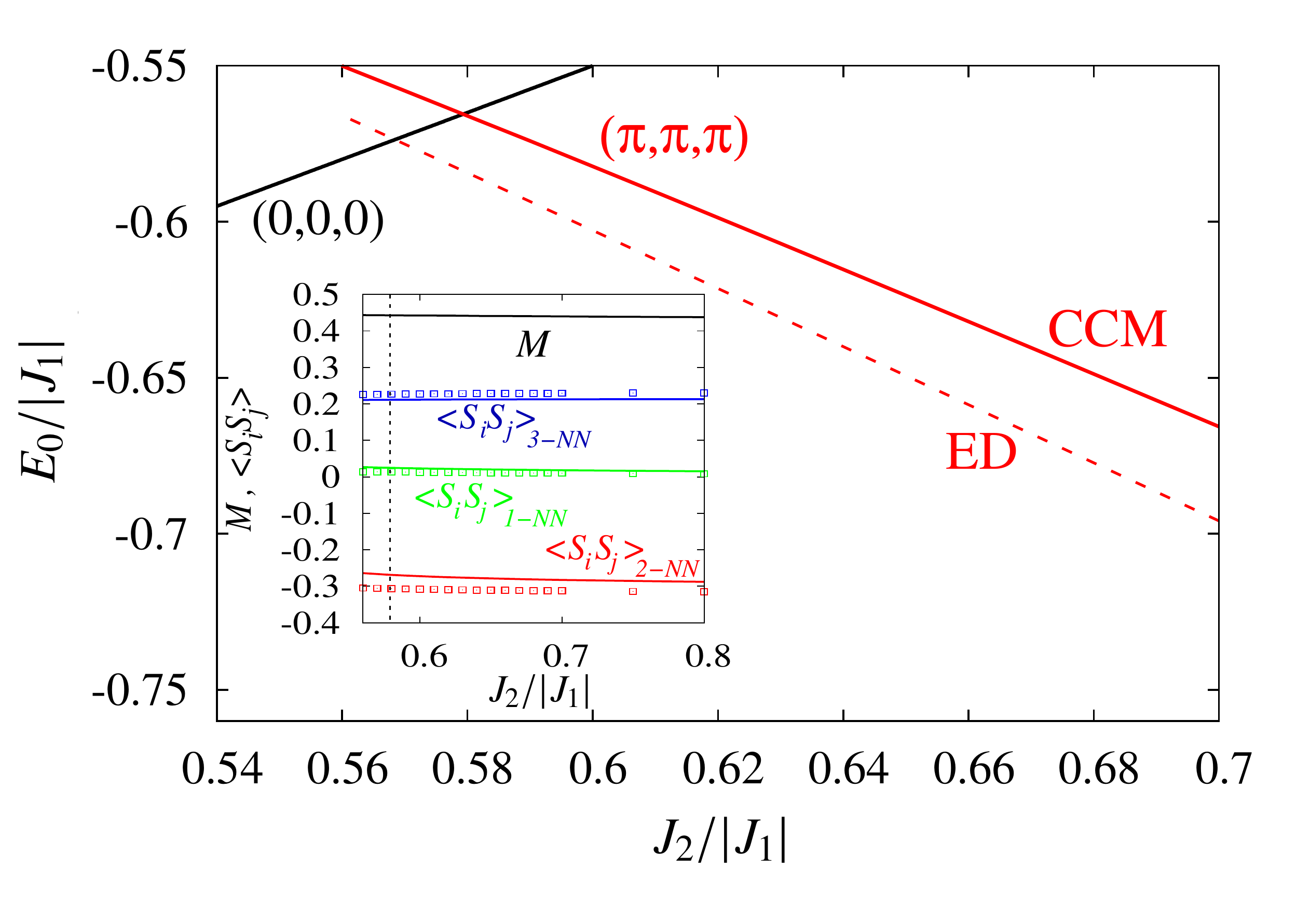}
\caption{Main panel:
Ground state energies per site for the $S=1/2$ $J_1$\textendash$J_2$ Heisenberg model on a BCC lattice with FM $J_{1}$. For the $(\pi,\pi,\pi)$ phase we use ED for a finite lattice of $N=36$ and the CCM for an infinite lattice. The crossing of the energies of the $(0,0,0)$ and $(\pi,\pi,\pi)$ states determines the first-order transition between both phases [see Table~\ref{tab:J2c}]. 
Inset: CCM result for the magnetic order parameter (sublattice magnetization) $M$ as well as ED and CCM results for the nearest-neighbor (1-NN), 
next-nearest-neighbor (2-NN) and third-nearest-neighbor (3-NN) equal-time spin-spin correlation functions of the $(\pi,\pi,\pi)$ phase 
(ED\textendash symbols, CCM\textendash lines).
The black vertical dotted indicates the first-order transition where  $(\pi,\pi,\pi)$ phase gives way for the ferromagnetic ground  state. 
}
\label{fig:ED-CCM}
\end{figure}

We now investigate the effects of quantum fluctuations on the classical phase diagram employing one-loop PFFRG. As found in Ref.~\cite{Rueck2018}, the one-loop formulation is mostly sufficient to correctly determine the phase boundaries for spin $S=1/2$. The more numerically-intensive two-loop scheme will only be applied for a select set of coupling parameters in order to calculate critical magnetic ordering temperatures more accurately, the results of which are presented in Sec.~\ref{sec:tc}. As described in Sec.~\ref{sec:frg}, at each point in parameter space we track the evolution of the susceptibility $\chi^{\Lambda}({\bf k})$ as a function of $\Lambda$ for \emph{all} $\mathbf{k}$ in the first Brillouin zone. The $\mathbf{k}$-vector which yields the dominant susceptibility at the point of breakdown of the RG flow then determines the nature of the magnetically ordered ground state. On the other hand, the absence of a breakdown in the limit $\Lambda\to0$ signals the absence of long-range dipolar magnetic order, and points to a paramagnetic ground state. We find that \emph{all} the magnetic orders in the classical phase diagram are to be found in the quantum phase diagram [Fig.~\ref{fig:cpd-qpd}(b) and Fig.~\ref{fig:cpd-qpd}(d)], and that no new types of long-range dipolar magnetic orders are stabilized by quantum fluctuations. In Fig.~\ref{fig:xivslam}, we show the representative RG flows for all the quantum phases with their momentum resolved susceptibility profiles shown in Fig.~\ref{fig:sus-FM} and Fig.~\ref{fig:sus-AFM}. 

For $S=1/2$, the most salient effect of quantum fluctuations is the appearance of a PM phase over an extended region in the $J_{2}$\textendash$J_{3}$ parameter space for FM as well as AF $J_{1}$ [see Fig.~\ref{fig:cpd-qpd}(b) and Fig.~\ref{fig:cpd-qpd}(d)]. The PM phase extends over a larger region in the case of FM $J_{1}$ as compared to AF $J_{1}$. It engulfs a substantial portion of parameter space occupied classically by the 2D and 3D incommensurate spiral orders, and to a lesser degree cuts into the classical domain of the $\mathbf{q}=(\pi,\pi,0)$ planar AF state. For FM $J_{1}$, we find that a tiny region of the classical FM phase is destabilized into a PM phase by quantum fluctuations. On the other hand, for AF $J_{1}$, the N\'eel order does not succumb at all to quantum fluctuations. Interestingly, we find that for FM as well as AF $J_{1}$ quantum fluctuations do not destabilize the $\mathbf{q}=(\pi,\pi,\pi)$ stripe AF order into a PM phase, and consequently, the PM phase does not occupy any portion of the parameter space which classically hosts the stripe AF. In Fig.~\ref{fig:sus-PM}, we present the momentum resolved susceptibilities for the PM phase. For completeness we have also considered larger spin magnitudes $S>1/2$, using a modification of the one-loop PFFRG as described in Ref.~\cite{Baez2017}. Interestingly, we find that already at $S=1$ the PM phase disappears entirely owing to the weakening of quantum fluctuations.

In the following, we investigate in more detail, the effects of quantum fluctuations for $S=1/2$ on the different magnetically ordered phases. Although, our principal method to investigate the quantum phase diagram is the PFFRG, for comparison  we also add here results using the exact diagonalization (ED) and the coupled-cluster method (CCM). Both methods have been used previously to study the $S=1/2$ BCC $J_{1}$\textendash$J_{2}$ model with AF $J_1$ \cite{Schmidt-2002,Farnell-2016}, but so far no results for FM $J_1$ are available. The ED is a well-established method, see, e.g.,~\cite{Schulz-1996,Schmidt-2002,richter2010-2d-j1j2}. Here, we use J.~Schulenburg's {\it spinpack}~\cite{spinpack1,spinpack2}. The CCM is a universal many-body method ~\cite{bishop1991overview,zeng1998efficient,Bishop98a_CCM,CCM_LNP2004} that has been successfully applied on frustrated quantum spin systems, see, e.g.~\cite{darradi2005coupled,darradi2008ground,Farnell2009high,CCMkagome2011,PhysRevB.91.014426,PhysRevB.95.134414,PhysRevB.98.224402}. We will give a brief illustration of the CCM in Appendix \ref{app:ccm}. Moreover, we mention that ED and CCM calculations used here  follow closely Ref.~\cite{Schmidt-2002} and Ref.~\cite{Farnell-2016}, respectively. The main ED and CCM results are summarized in Fig.~\ref{fig:ED-CCM}. 

\floatsetup[table]{capposition=bottom}
\begin{table*}
\setlength{\tabcolsep}{8pt}
\setlength\extrarowheight{2pt}
\centering
\begin{tabular}{llllllllll}
\hline\hline
 & Phase I & Phase II & Method & $J_{2}^{c}/|J_1|$ \\
 \hline
 \multirow{5}{*}{$J_1$\textendash Ferromagnetic}& \multirow{5}{*}{$(0,0,0)$} &  \multirow{5}{*}{$(\pi,\pi,\pi)$}&{PFFRG$^{*}$} & $0.56(2)$    \\
& & &Exact Diagonalization$^{*}$ & $0.568$      \\
& & &Coupled Cluster Method$^{*}$ & $0.579$      \\
& & &Rotation-invariant Green's function method~\cite{Mueller-2015} & $0.68$      \\
& & &Random phase approximation~\cite{Raza-1964} & $0.6799$  \\
 \hline
 \multirow{6}{*}{$J_1$\textendash Antiferromagnetic}& \multirow{6}{*}{$(2\pi,0,0)$} &  \multirow{6}{*}{$(\pi,\pi,\pi)$}&{PFFRG$^{*}$} & $0.70(2)$    \\
& & &Coupled Cluster Method~\cite{Farnell-2016} & $0.704$      \\
& & &Exact Diagonalization~\cite{Schmidt-2002} & $0.7$  \\
& & &Non-linear spin-wave theory~\cite{Majumdar-2009} & $0.705$      \\
& & &Random phase approximation~\cite{Pantic-2014} & $0.72$  \\
& & &Linked Cluster Series expansions~\cite{Oitmaa-2004} & $0.705(5)$  \\
 \hline\hline
\end{tabular}

\caption{The critical value $J_{2}^{c}/|J_1|$ of the transition between the FM/N\'eel and the stripe order obtained from PFFRG and compared to different methods for the $S=1/2$ $J_1$\textendash$J_2$ Heisenberg model on the BCC lattice with $J_3=0$. The results marked with an asterisk are from the present study.}
\label{tab:J2c}
\end{table*}

\begin{figure*}[htp]
\includegraphics[width=1\columnwidth]{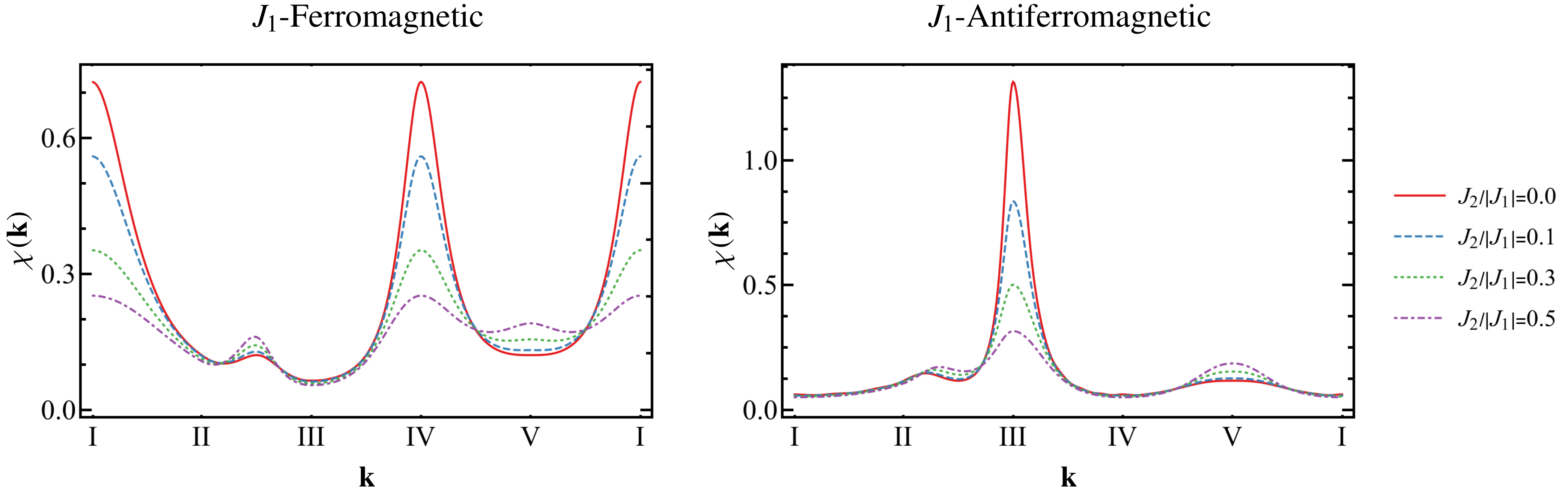}
\caption{The susceptibility $\chi(\mathbf{k})$ as a function of the wave vector $\mathbf{k}$ for different values of $J_2$, with both FM (left panel) and AF (right panel) $J_1$, along the high symmetry lines in the momentum space. The characteristic points of our choice are: $\text{I}= (0, 0, 0)$, $\text{II}= (0, 0, \pi)$, $\text{III}= (2\pi, 2\pi, 2\pi)$, $\text{IV}= (2\pi, 2\pi, 0)$, $\text{V}= (\pi, \pi, \pi)$. The susceptibility is plotted for $\Lambda/|J_{1}|\simeq4/\pi$ which corresponds to a temperature $T/|J_{1}|\simeq2$. Note that the ordering wave vector $\mathbf{q}=(0,0,0)$ is equivalent to $\mathbf{q}=(2\pi,2\pi,0)$. Similarly, the ordering wave vector $\mathbf{q}=(2\pi,0,0)$ is equivalent to $\mathbf{q}=(2\pi,2\pi,2\pi)$.}\label{fig:sus-PM1}
\end{figure*}

Firstly, the only impact of quantum fluctuations on the stripe AF state is to shift its phase boundaries compared to the classical ones. In particular, we find that for large enough $J_{2}$ (i) the stripe AF replaces the classical incommensurate 3D spirals, whose existence is thus reduced to a tiny sliver in the $J_{2}$\textendash$J_{3}$ plane and (ii) the stripe AF cuts into the classical domain of the planar AF with which it now shares a phase boundary hitherto absent in the classical phase diagram. This is similar to the findings on the square lattice wherein quantum fluctuations are found to favor the $(\pi,\pi)$ state over the $(\pi,0)$ state~\cite{Danu-2016}. In contrast, quantum fluctuations act differently on the other phase boundary of the stripe AF with the FM or N\'eel orders depending on whether $J_{1}$ is FM or AF, respectively. For FM $J_{1}$, we find that the phase boundary shifts to a smaller value of $J_{2}$, whereas for AF $J_{1}$ the phase boundary is shifted to a larger value as also observed on the simple cubic lattice~\cite{Iqbal-2016a}. In particular, there is \emph{no} intermediate PM phase in between the FM/N\'eel and stripe AF orders in the $J_{1}$\textendash$J_{2}$ model in agreement with previous studies~\cite{Raza-1964,Mueller-2015,Farnell-2016,Schmidt-2002,Majumdar-2009,Pantic-2014,Oitmaa-2004}. This is in contrast to the findings on the square lattice, and can be attributed to the diminished quantum fluctuations in 3D. In Table~\ref{tab:J2c}, we provide numerical estimates of the phase boundary for $J_{3}=0$, i.e., along the $J_{1}$\textendash$J_{2}$ line obtained by PFFRG and other numerical approaches. The observation that for FM $J_1$ the stripe AF order extends at the expense of FM order follows from the fact that quantum fluctuations, in general, do not alter the FM state (including its ground state energy) as its an eigenstate of the Heisenberg exchange Hamiltonian and thus free of macroscopic zero-point vibrations~\cite{Nagaev-1984,Kaganov-1987}, however, act on AF orders, e.g., by lowering their ground state energies. Hence, compared to the classical case, phase boundaries between FM and AF orders are typically shifted towards the FM side. One further observes that the shift in the phase boundary from the classical value of $J_{2}^{c}/|J_{1}|=2/3$ is stronger for FM $J_{1}$ compared to AF $J_{1}$. We mention that our PFFRG findings for the BCC $J_{1}$\textendash$J_{2}$ model are in excellent agreement with ED and CCM results, cf. Table~\ref{tab:J2c}, where we provide a comparison of the critical values $J_{2}^{c}/|J_1|$ of the transition between the FM/N\'eel and the stripe orders. Moreover, the ED and CCM results for the spin-spin correlation functions and the order parameter shown in Fig.~\ref{fig:ED-CCM} clearly demonstrate an absence of an intermediate quantum paramagnetic phase, cf. the PFFRG phase diagram in Fig.~\ref{fig:cpd-qpd}(b).

A plot of the susceptibility along a path in reciprocal space for different $J_{2}/|J_{1}|$ within the FM and N\'eel ordered phases is presented in Fig.~\ref{fig:sus-PM1}. The maxima of the susceptibility at the magnetic wave vectors of the respective orders are seen to be clearly resolved. The frustrating effect of a $J_{2}$ coupling on the FM/N\'eel orders leads to a reduction in the dominant susceptibility peaks and to the development of a peak at the incipient stripe AF order at $\mathbf{q}=(\pi,\pi,\pi)$, similar to the findings by high temperature series expansion~\cite{Richter-2015}.  

\begin{figure*}
\includegraphics[width=1.0\columnwidth]{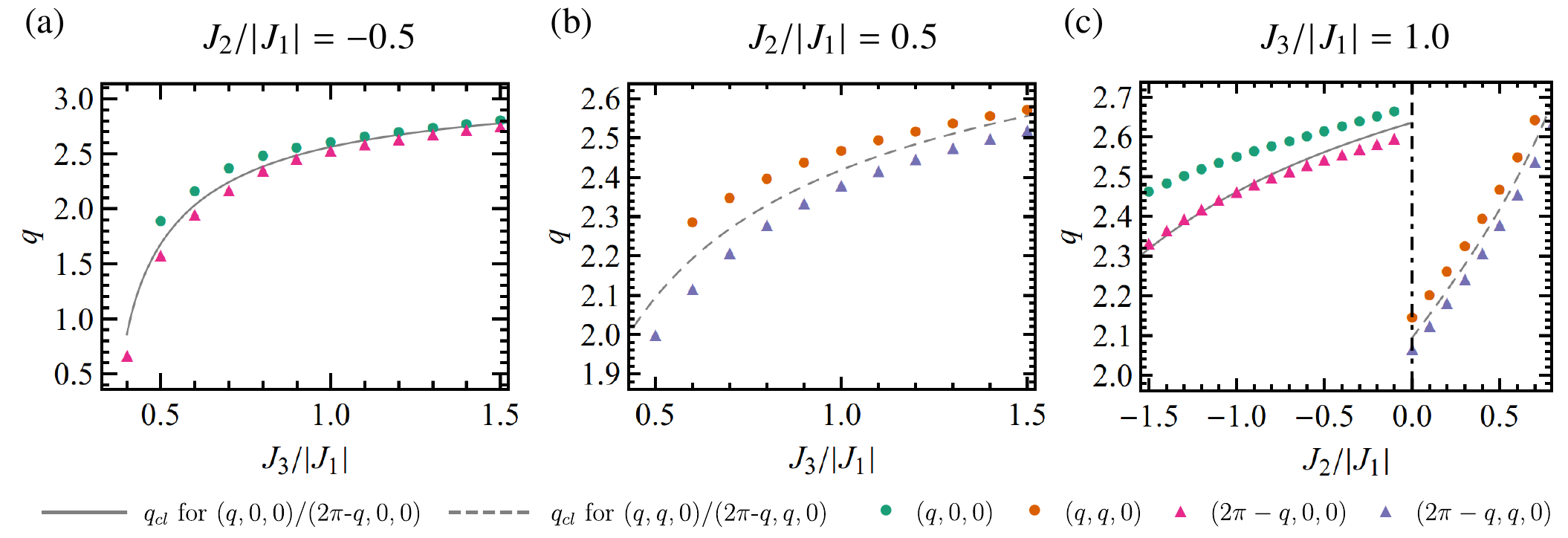}
\caption{Variation of $q$ for the incommensurate spiral phases $\q00$, $\qq0$, $\tpmq00$, and $\tpmqq0$ along three different cuts in the $J_{2}$\textendash$J_{3}$ plane of the Heisenberg model on the BCC lattice. Circles (triangles) denote the case of FM (AF) $J_1$. The corresponding classical values are shown by gray lines.}\label{fig:shift}
\end{figure*}

Another phase boundary which is significantly shifted by quantum fluctuations is the one between the 2D spiral and planar AF orders [see Fig.~\ref{fig:cpd-qpd}]. For FM as well as AF $J_{1}$, the classical phase boundary at $J_{2}/|J_{1}|=1$ is shifted to a smaller value, implying that quantum effects enhance the stability of the planar AF state. In particular, for FM $J_{1}$ the phase boundary shifts to $J_{2}/|J_{1}|\approx0.82$, while for AF $J_{1}$ the shift is comparatively weaker, and the phase boundary is found to be located at $J_{2}/|J_{1}|\approx0.93$. This strong effect of quantum fluctuations can be explained by the fact that the $\mathbf{q}=(\pi,\pi,0)$ planar AF has a \emph{continuous} $8$\textendash dimensional classical ground state degeneracy whereas the 2D spiral only features an $8$\textendash fold \emph{discrete} degeneracy. Consequently, quantum fluctuations play a more prominent role on top of the classical planar AF state in comparison to the 2D spiral. 

Similar to the shifts observed in the phase boundary between the FM/N\'eel and stripe AF order, we find that the boundaries between FM/N\'eel orders to the 1D and 3D spiral phases behave differently depending on whether $J_{1}$ is FM or AF. For FM $J_{1}$, the 1D and 3D spirals enhance their domain of stability beyond the classically allowed region of their existence, and thus the domain of the FM phase shrinks compared to the classical one. In contrast for AF $J_{1}$, the N\'eel phase extends into these spiral orders but to a lesser extent. In total, the domain of existence of the FM order is significantly reduced by quantum fluctuations, whereas for the N\'eel phase it is enhanced, compared to the classical phase diagram. Finally, two phase boundaries remain unaffected by quantum fluctuations, namely, the 2D to 3D spirals, and the one between 1D and 2D spirals. 
 
Now, we discuss the impact of quantum fluctuations within the incommensurate spiral orders which primarily amounts to a shift of the value of the pitch vectors. Indeed, our PFFRG analysis shows that for FM $J_{1}$ the shift is such that the 1D spiral pitch vector is shifted towards that of the N\'eel state. At a fixed $J_{2}$, this effect is stronger for small $J_{3}$ and appears to decrease with increasing $J_{3}$ [Fig.~\ref{fig:shift}(a)]. At a fixed $J_{3}$, this shift of $\mathbf{q}$ towards the N\'eel pitch vector increases with increasing FM $J_{2}$ [Fig.~\ref{fig:shift}(c)]. Note that quantum fluctuations seem to act counterintuitively here since with increasing strength of the FM $J_{2}$ coupling one would expect the FM state to become increasingly favorable. Interestingly, in the case of AF $J_1$ the 1D spiral pitch vectors shift in the opposite direction, i.e., they migrate towards the FM state [see Fig.~\ref{fig:shift}(a)], except for large FM $J_2$ where essentially no shift is observed [see Fig.~\ref{fig:shift}(c)]. For the 2D spirals, we observe that when $J_{1}$ is FM, the shift of the pitch vector is towards that of the planar AF order, and the magnitude of the shift decreases with increasing $J_{3}$ [see Fig.~\ref{fig:shift}(b)]. At fixed $J_{3}$, the magnitude of the shift remains essentially constant with varying $J_{2}$ [see Fig.~\ref{fig:shift}(c)]. This finding is consistent with the fact that the phase boundary of the planar AF state shifts to a smaller value of $J_{2}$. However, for AF $J_{1}$ we find that the 2D spiral pitch vector is shifted towards that of the N\'eel state, and upon varying $J_{2}$ and $J_{3}$ the magnitude of the shift remains essentially constant [see Fig.~\ref{fig:shift}(b) and Fig.~\ref{fig:shift}(c)]. We do not show the changes of the $\bf q$ vector for the $\qqq$ spiral since it covers only a tiny sliver in the quantum phase diagram. 

\subsection{N\'eel and Curie temperatures}\label{sec:tc}

\floatsetup[table]{capposition=bottom}
\begin{table*}
\setlength{\tabcolsep}{6pt}
\setlength\extrarowheight{2pt}
\centering
\begin{tabular}{llllllllll}
\hline\hline
 & Method & $\frac{J_{2}}{|J_{1}|}=0$ & $\frac{J_2}{|J_1|}=0.1$ & $\frac{J_2}{|J_1|}=0.2$ & $\frac{J_2}{|J_1|}=0.3$ &$\frac{J_2}{|J_1|}=0.4$ & $\frac{J_2}{|J_1|}=0.5$ &$\frac{J_2}{|J_1|}=0.6$ \\
 \hline
 \multirow{7}{*}{$J_1$\textendash FM} & {PFFRG (one loop)$^{*}$} & $1.45(1)$ & $1.33(1)$ & $1.18(1)$ & $1.04(1)$ & $0.88(1)$ & $0.66(1)$     \\
&{PFFRG (two loop)$^{*}$} & $1.37(1)$ & $1.26(3)$ & $1.11(1)$ & $0.97(1)$ & $0.84(3)$ & $0.62(1)$     \\
&{QMC$^{*}$} & $1.260(1)$ & $$ & $$ & $$ & $$ & $$ \\
&{HTE [$\chi_{\rm u}$]}~\cite{Rushbrooke,Oitmaa1996,Zheng-2004} & $1.2602(5)$ & $ $ & $ $ & $ $ & $ $  \\
&{HTE [$\chi_{\rm u}^{\rm quot}$]}~\cite{Mueller-2015} & $1.253(8)$ & $1.125(15)$ & $ $ & $1.06(32)$ & $ $  \\
&{HTE [$\chi_{\rm u}^{\rm DA}$]}~\cite{Mueller-2015} & $1.268(8)$ & $1.16(3)$ & $ $ & $0.89(7)$ & $ $  \\
&{HTE [${S({\bm Q}})$]}~\cite{Richter-2015} & $1.273(8)$ & $1.17(3)$ & $1.09(10)$ & $0.89(7)$ & $0.72(11)$ & $0.45(18)$ \\
&{GFA}~\cite{Mueller-2015} & $1.359$ & $1.247$ & $1.136$ & $1.022$  & $0.903$ & $0.771$\\
 \hline
\multirow{7}{*}{$J_1$\textendash AF} & PFFRG (one loop)$^{*}$ & $1.63(1)$ & $1.52(1)$ & $1.36(1)$ & $1.26(2)$ & $1.10(1)$ & $0.90(1)$ & $0.71(1)$  \\
 & PFFRG (two loop)$^{*}$ & $1.50(1)$ & $1.42(1)$ & $1.32(1)$ & $1.16(1)$ & $1.06(1)$ & $0.89(1)$ & $0.75(1)$  \\
 &{QMC$^{*}$} & $1.377(2)$ & $$ & $$ & $$ & $$ \\
 & HTE [$\chi_{\rm s}$]~\cite{Zheng-2004}& $1.376(4)$ & $$ & $$ & $$ & $$ & $$ & $$  \\
 & HTE $[{S({\bm Q})}]$~\cite{Richter-2015}& $1.50(8)$ & $1.36(10)$ & $1.26(13)$ & $1.09(13)$ & $0.96(7)$ & $0.75(6)$ & $0.61(10)$  \\
 & HTE {[$\chi_{\rm s}$]~\cite{Oitmaa-2004}} & $1.38(2)$ & $1.26(2)$ & $1.13(1)$ & $1.00(2)$ & $0.86(2)$ & $0.71(2)$ & $0.57(3)$ \\
 & GFA~\cite{Richter-2015} &$1.530$ & $1.369$ & $1.195$ & $1.004$ & $0.786$ & $0.520$ & $$   \\
 \hline\hline
\end{tabular}

\caption{Critical temperatures $T_{c}/|J_1|$ of the $S=1/2$ $J_1$\textendash$J_2$ BCC Heisenberg model for various values of the frustrating AF $J_{2}$ coupling, obtained from different methods for both FM as well as AF $J_{1}$. The results marked with an asterisk are from the present study. For plotted data see Fig.~\ref{fig:TCTN}.}
\label{tab:TCTN}
\end{table*}

\floatsetup[table]{capposition=top}
\begin{table*}
\setlength{\tabcolsep}{8pt}
\centering
\begin{tabular}{llllll}
 \hline \hline
      \multicolumn{1}{l}{Lattice}
    & \multicolumn{1}{c}{$T_{C}^{S=1/2}/[|J_{1}|S(S+1)]$}
    & \multicolumn{1}{c}{$T_{N}^{S=1/2}/[J_{1}S(S+1)]$}
    & \multicolumn{1}{c}{$T_{C/N}^{S\to\infty}/[|J_{1}|S(S+1)]$}
    & \multicolumn{1}{c}{$\frac{T_{C}^{S=1/2}}{T_{C/N}^{S\to\infty}}$} 
    & \multicolumn{1}{c}{$\frac{T_{N}^{S=1/2}}{T_{C/N}^{S\to\infty}}$}  \\ \hline
       
\multirow{1}{*}{BCC} & $1.680(1)$ & $1.836(3)$ & $\num{2.054241(52)}$~\cite{Chen-1993} & $0.818(1)$ & $0.894(3)$    \\

\multirow{1}{*}{SC} & $1.119(1)$~\cite{Troyer-2004,Wessel-2010} & $1.261(1)$~\cite{Sandvik-1998} & $\num{1.442929(77)}$~\cite{Chen-1993} & $0.776(1)$ & $0.874(1)$ \\ \hline \hline

\end{tabular}
\caption{For the nearest-neighbor Heisenberg FM and the AF on the BCC and simple cubic (SC) lattices, we provide for $S=1/2$, the Curie $T_{C}^{S=1/2}/[|J_{1}|S(S+1)]$ temperature (column 2) and the N\'eel temperature $T_{N}^{S=1/2}/[J_{1}S(S+1)]$ (column 3) as obtained from Quantum Monte Carlo. We also provide the ordering temperature $T_{C/N}^{S\to\infty}/[|J_{1}|S(S+1)]$ for the corresponding classical Heisenberg model as obtained from Classical Monte Carlo. In the classical model, the equality of the ordering temperatures for FM exchange and AF exchange is a consequence of the fact that the free energy is an even function of the coupling $J_{1}$~\cite{Zheng-2004}.}
\label{tab:t-neel}
\end{table*}

The magnetic ordering temperature, i.e., the Curie temperature ($T_{C}$) for a FM and the N\'eel temperature ($T_{N}$) for an AF ordered state is one of the fundamental thermodynamic quantities which serves as a measure of the degree of frustration. The (numerically exact) quantum Monte Carlo (QMC) method can be employed to calculate this $T_{C}$ and $T_{N}$ in the non-frustrated region of parameter space. However, in the frustrated regime, one must resort to approximate numerical approaches to obtain estimates of $T_{C}$ and $T_{N}$. Here, we employ one- and two-loop PFFRG to estimate the ordering temperatures for non-frustrated and frustrated coupling parameters of the $J_{1}$\textendash$J_{2}$ BCC Heisenberg model. As observed in Ref.~\cite{Rueck2018}, the one-loop PFFRG is less converged when estimating critical ordering temperatures as compared to determining phase boundaries. We, therefore, carry out our calculations in both one-loop \emph{and} two-loop formulation. Furthermore, we compare these estimates to those obtained by high temperature expansion (HTE), and Green's function methods in previous studies which also serves and a benchmark test for the performance of the PFFRG. Additionally, we carry out QMC calculations for the nearest-neighbor FM and AF couplings only, and obtain estimates of $T_{C}$ and $T_{N}$ by a finite-size scaling analysis of the renormalization-group invariant quantities Binder ratio and the ratio $\xi/L$ of the second-moment correlation length $\xi$ over the lattice size $L$; see Appendix A of Ref.~\cite{PTHAH-14} for a discussion on the definition of $\xi$. More details on the QMC simulations and the analysis are reported in Appendix~\ref{app:qmc}.

\begin{figure*}
\includegraphics[width=1\textwidth]{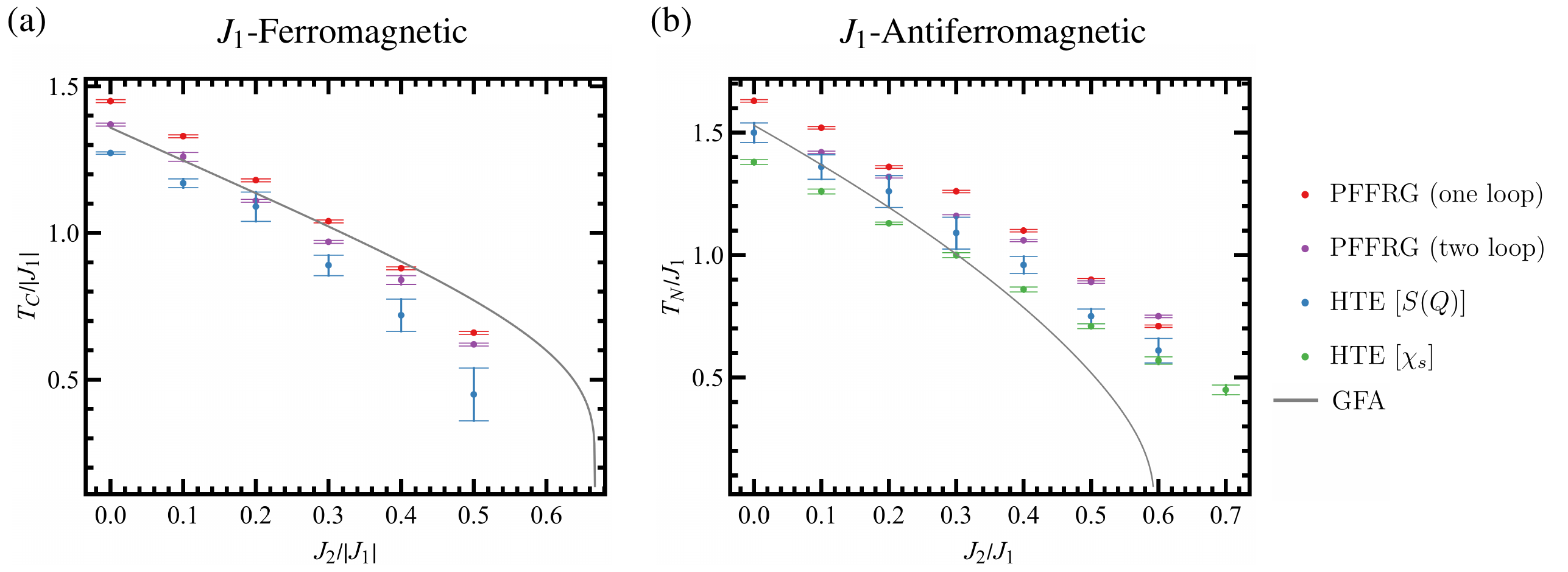}
\caption{Critical temperatures $T_{c}$ of the $J_1$\textendash$J_2$ BCC Heisenberg model as a function of the frustrating AF $J_{2}$ coupling. Left panel: Curie temperatures $T_C/|J_{1}|$ for $J_1$-FM. Right panel: N\'eel temperatures $T_N/J_{1}$ for $J_1$-AF. Please refer to Table~\ref{tab:TCTN} for numerical values.}\label{fig:TCTN}
\end{figure*}

We start discussing the $S=1/2$ nearest neighbor only model with FM and AF $J_1$ interaction. Both systems are unfrustrated such that they are amenable to a QMC calculation. Interestingly, our QMC results show that the N\'eel and the Curie temperatures are unequal [see Table~\ref{tab:TCTN}], with $T_{N}$ being greater than $T_{C}$ by about $9\%$, in agreement with the findings from HTE and Green's function methods~\cite{Rushbrooke-1963,Rushbrooke,Oitmaa-2004,Junger-2009,Richter-2015} (see Table~\ref{tab:TCTN}). Indeed, it is known to be a general feature of a finite spin-$S$ Heisenberg models on bipartite lattices with nonfrustrating interactions that the N\'eel and Curie temperatures are unequal with $T_{N}>T_{C}$~\cite{Zheng-2004}. This difference between $T_{C}$ and $T_{N}$ of $\sim9\%$ is also reflected in our two-loop PFFRG data but is slightly overestimated on the one-loop level where the difference is found to be $\sim12\%$. Concerning absolute values of the ordering temperatures, our one-loop and two-loop results both slightly overestimate $T_{N}$ and $T_{C}$. By extending the PFFRG from one-loop to two-loop the accuracy of the results becomes significantly better, particularly, the errors of the one-loop critical ordering temperatures are approximately halved in the two-loop results. One may therefore expect that even higher loop orders might give very accurate estimates. We leave such an analysis for future studies. Finally, in Table~\ref{tab:t-neel}, we compare our results against those for the simple cubic lattice, and also compare the $S=1/2$ ordering temperatures against the ones for the classical model to obtain the reduction due to quantum fluctuations. As expected, the ordering temperatures for the BCC lattice are larger compared to those for the SC lattice due to its higher coordination number. For the same reason, the reduction in the critical temperature for $S=1/2$ with respect to the classical value is lesser for the BCC lattice in comparison to the SC lattice.  

In the presence of a frustrating $J_{2}$ interaction there is a significant reduction in both $T_{C}$ and $T_{N}$, which are found to decrease monotonically with increasing $J_{2}$ [see Fig.~\ref{fig:TCTN}], and on approaching the transition point the ordering temperatures have a sharp drop. Our PFFRG data shows that the inequality $T_{N}>T_{C}$ remains valid up till the transition point into the stripe AF order in agreement with HTE data~\cite{Oitmaa-2004,Zheng-2004,Mueller-2015,Richter-2015}, but in contrast to results from Green's function approach~\cite{Mueller-2015,Richter-2015} [see Table~\ref{tab:TCTN}]. Again, we see that the ordering temperatures from PFFRG are slightly larger than those obtained by HTE where two-loop PFFRG mostly gives better estimates. Overall, these results imply that PFFRG (particularly the two-loop formulation) correctly captures the relative behavior of ordering temperatures. The absolute values, however, might still be subject to errors of a few percent which are possibly reduced within higher-loop schemes.  

\section{Summary and Outlook}\label{sec:conclusion}

We have shown that frustrating the ferromagnetic and N\'eel antiferromagnetic orders of Heisenberg spins on a three-dimensional bipartite body-centered-cubic lattice by competing interactions up to third neighbors leads to the appearance of a rich variety of helimagnetic and collinear spin structures at the classical level. In the extreme quantum limit of $S=1/2$, our PFFRG analysis shows that the most salient feature of quantum fluctuations is the realization of an extended region of parameter space displaying quantum paramagnetic behavior. The classical phase boundaries are also found to be strongly renormalized by quantum effects, and helimagnetic pitch vectors undergo significant shifts. In total, we find that quantum effects are stronger in the case of a ferromagnetic nearest-neighbor coupling compared to an antiferromagnetic one. We have also estimated the Curie and N\'eel temperatures from PFFRG, and compared our results to those from quantum Monte Carlo for unfrustrated case of nearest-neighbor only antiferromagnetic and ferromagnetic models, and with available high temperature expansion data in the frustrated regime of the $J_{1}$\textendash$J_{2}$ model. We obtain good agreement with Quantum Monte Carlo for the pure nearest-neighbor Heisenberg ferromagnet and N\'eel antiferromagnet and reproduce qualitative trends for frustrating couplings. However, we observe that in general the PFFRG overestimates the ordering temperatures which is partially cured by employing two-loop PFFRG.

As a future study, it will be interesting to investigate the finite-temperature classical phase diagram of the $J_{1}$\textendash$J_{2}$\textendash$J_{3}$ model, including its critical properties and nature of phase transitions, which has traditionally largely focussed on the Ising model~\cite{Banavar-1979,Velgakis-1983,Azaria-1989,Murtazaev-2015,Murtazaev-2017}, however, recent attempts have been made at the Heisenberg model for a given parameter value~\cite{Murtazaev-2018,Ramazanov-2017}. The role of disorder in determining the stability of the realized phases is another important issue worth investigating. It has been pointed out in Ref.~\cite{Attig-2017} that \emph{if} one restricts the second nearest-neighbor coupling $J_{2}^{*}$ to be defined by bond-distance instead of geometrical distance (as in the current paper), then the classical Heisenberg $J_{1}$\textendash$J_{2}^{*}$ antiferromagnet on the BCC lattice hosts spin spiral surfaces analogous to the $J_{1}$\textendash$J_{2}$ model on the diamond lattice~\cite{bergmann07}. It will be interesting to investigate the selection effects on the spiral surface due to quantum fluctuations as a function of the frustration ratio $J_{2}^{*}/J_{1}$ and spin-$S$, and in particular, examine the possibility of realizing a spiral spin liquid. Our finding of extended domains characterized by an absence of long-range dipolar magnetic order in the $S=1/2$ model lays the avenue for future numerical investigations aiming to identify the nature of the nonmagnetic phase which could potentially be host to a plethora of exotic nonmagnetic phases such as quantum spin liquids, valence-bond-crystals, and lattice-nematics \emph{or} feature quadrupolar ordered phases, i.e., spin-nematic orders~\cite{Andreev-1984}. Indeed, in the $S=1/2$ $J_{1}$\textendash$J_{2}$\textendash$J_{3}$ square lattice Heisenberg model, these orders were found to be stabilized~\cite{Iqbal-2016}. The question of the microscopic identification of the nature of the nonmagnetic phase can be addressed within the PFFRG framework itself by combining it with a self-consistent Fock-like mean-field scheme to calculate low-energy effective theories for emergent spinon excitations in $S=1/2$ systems as has been recently achieved on the square and kagome lattices~\cite{Hering-2019}. Within this scheme, the effective spin interactions obtained from PFFRG, i.e., the two-particle vertices, act as an input for the Fock equation yielding a self-consistent approach to calculate the spinon band structures beyond a mean field treatment. However, the precise forms of such free spinon Ans\"atze are given by a projective symmetry group classification~\cite{Wen-2002}, and it will be useful to carry out a classification of the symmetry allowed mean-field quantum spin liquid and nematic states on the BCC lattice. These Ans\"atze would also then serve as the basis for Gutzwiller projected variational wave-function studies employing Monte Carlo methods~\cite{Iqbal-2011b,Hu-2013,Iqbal-2013,Iqbal-2016a}.

{\it Acknowledgments}. 
T.M., P.G., and Y.I. thank R. Ganesh and Arnab Sen for helpful discussions. J. Richter  thanks O. G\"otze for providing the CCM scriptfiles for the AF
 BCC model.
 We gratefully acknowledge the Gauss Centre for Supercomputing e.V. for funding this project by providing computing time on the GCS Supercomputer SuperMUC at Leibniz Supercomputing Centre (LRZ). The work in W\"urzburg
was supported by the DFG through DFG-SFB 1170 tocotronics (project B04) and the W\"urzburg-Dresden Cluster of Excellence on Complexity and Topology in Quantum Matter DFG-EXC 2147/1  \textit{ct.qmat} (project-id 39085490). 
F.P.T. thanks the German Research Foundation (DFG) through Grant No. AS120/13-1 of the FOR 1807. Y.I. acknowledges the kind hospitality of the Helmholtz-Zentrum f\"ur Materialien und Energie, Berlin, Germany for the period May--July 2018, where part of this work was accomplished. P.G., Y.I., and T.M. acknowledge the interactions and kind hospitality at the International Centre for Theoretical Sciences (ICTS), Bengaluru, India from 29th November till 7 December, 2018 during ``The $2$nd Asia Pacific Workshop on Quantum Magnetism''. 


%

\appendix

\section{Brief illustration of the coupled-cluster method (CCM)}\label{app:ccm}

We illustrate here only some features of
the CCM relevant for the results shown in Fig.~\ref{fig:ED-CCM}. At that we follow the lines given in
Ref.~\cite{Farnell-2016}, where the CCM was applied to the 
$J_1$\textendash$J_2$ BCC model with
AF $J_{1}$.
For more general information on the methodology of the CCM, see, e.g.,
Refs.~
\cite{bishop1991overview,zeng1998efficient,Bishop98a_CCM,CCM_LNP2004}.
We first mention that the CCM yields results directly in 
the thermodynamic limit $N\to\infty$. 
First we choose 
 a normalized reference or model  state
$|\Phi\rangle$ that is here the classical $(\pi,\pi,\pi)$-state, see
Fig.~\ref{fig:mag-structure}.
Second we perform a rotation of the local axes of each of 
the spins such that all spins in the model state align along the
negative $z$ axis.
In this new set of local spin coordinates 
we define a complete set of 
mutually commuting multispin
creation operators $C_I^+ \equiv (C^{-}_{I})^{\dagger}$ related to this
model 
state:
$|{\Phi}\rangle = |\downarrow\downarrow\downarrow\cdots\rangle ; \mbox{ }
C_I^+ 
= \hat{ S}_{n}^+ \, , \, \hat{ S}_{n}^+\hat{ S}_{m}^+ \, , \, \hat{ S}_{n}^+\hat{ S}_{m}^+{
s}_{k}^+ \, , \, \ldots \; ,
$
$\hat{ S}^{+}_{n} \equiv \hat{ S}^{x}_{n} + i\hat{ S}^{y}_{n}$, where the spin operators
$\hat{ S}^{x}_{n}$ and $\hat{ S}^{y}_{n}$  are defined 
in the local rotated coordinate frames, and the indices $n,m,k,\ldots$ denote arbitrary lattice
sites.
The  CCM parameterizations of 
the ket 
and bra GS eigenvectors
$|\Psi\rangle$ 
and $\langle\tilde{\Psi}|$ 
of the spin system 
are given  by
$|\Psi\rangle=e^S|\Phi\rangle \; , \mbox{ } S=\sum_{I\neq 0}a_IC_I^+ \; ; \;
$
$\langle \tilde{ \Psi}|=\langle \Phi |\tilde{S}e^{-S} \; , \mbox{ } \tilde{S}=1+
\sum_{I\neq 0}\tilde{a}_IC_I^{-}$.
The coefficients
$a_I$ and $\tilde{a}_I$ 
contain the CCM correlation operators, $S$ and $\tilde{S}$.
They
are determined by the ket-state 
and bra-state
equations
$\langle\Phi|C_I^-e^{-S}He^S|\Phi\rangle = 0 \; ; \; 
\langle\Phi|{\tilde S}e^{-S}[H, C_I^+]e^S|\Phi\rangle = 0 \;  ; \; \forall
I\neq 0.$
Each of these 
equations labeled by a configuration index $I$,
corresponds to a certain configuration of lattice sites
$n,m,k,\dots\;$.
Using the Schr\"odinger equation, $H|\Psi\rangle=E_0|\Psi\rangle$, we can write
the ground-state energy per site as $E_0=\langle\Phi|e^{-S}He^S|\Phi\rangle$.
The magnetic order parameter (sublattice magnetization) is given
by $ M = -\frac{1}{N} \sum_{i=1}^N \langle\tilde\Psi|\hat{ S}_i^z|\Psi\rangle$, where
$\hat{ S}_i^z$
is expressed in the transformed coordinate system, and $N(\rightarrow \infty)$ is the number of lattice sites. 
In order 
to truncate the expansions of $S$ 
and $\tilde S$ we use the well established LSUB$n$ approximation scheme, cf., e.g., Refs.~\cite{zeng1998efficient,CCM_LNP2004,darradi2005coupled,darradi2008ground,Farnell2009high,CCMkagome2011,PhysRevB.91.014426,PhysRevB.95.134414,PhysRevB.98.224402}.  
 In the LSUB$n$ scheme 
all multi-spin correlations over distinct
locales on the lattice defined by $n$ or fewer contiguous
sites are retained. 
Using an efficient 
parallelized CCM code \cite{cccm} we solve the CCM equations up
to LSUB8.
The maximum number of ket-state equations considered here 
is  128267.
For the considered $(\pi,\pi,\pi)$ state of $J_1$\textendash$J_2$ BCC model with
FM $J_{1}$ we find that the LSUB$n$ data rapidly converge to the $n \to \infty$ limit.
Thus, the difference  betweeen the LSUB6 and LSUB8 CCM ground state energies
$E_0$ (order parameter $M$)
is less than 0.1\% (1\%).
Therefore, the LSUB8 data used for the CCM curves shown in
Fig.~\ref{fig:ED-CCM} practically may stand for the converged $n \to \infty$
data.  

\section{Quantum Monte Carlo analysis}\label{app:qmc}

\begin{figure*}
  \centering
  \includegraphics[width=\linewidth]{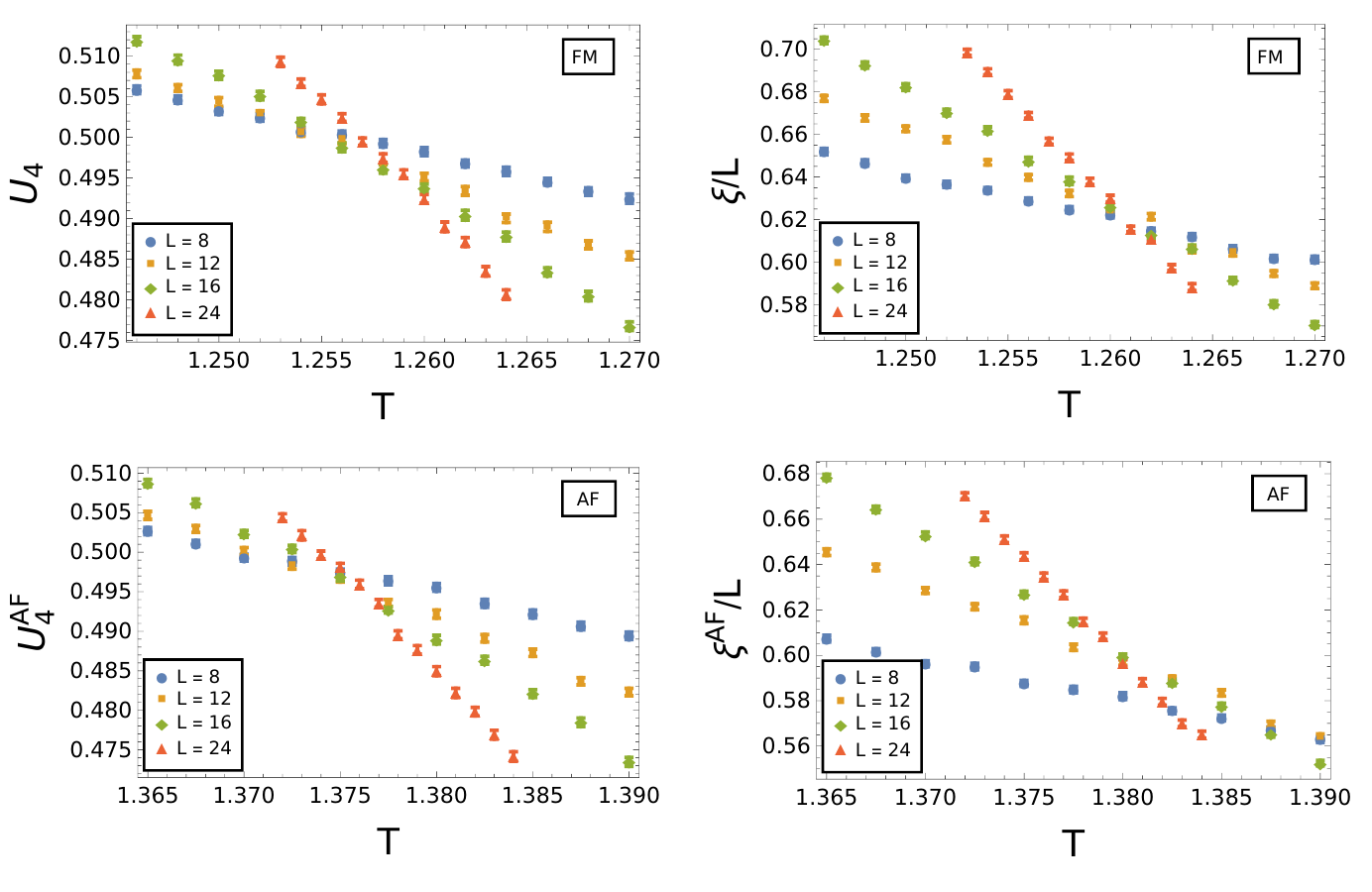}
  \caption{QMC estimates for the Binder ratio $U_4$ and the ratio of the finite-size correlation length over the lattice size $\xi/L$ for a FM Heisenberg model on the BCC lattice, and the corresponding quantities for the AF model, $U_4^{\rm AF}$ and $\xi^{\rm AF}/L$.}
  \label{fig:rg}
\end{figure*}

\newcolumntype{v}{D{.}{.}{1.7}}
\newcolumntype{f}{D{/}{/}{4.2}}
\begin{table*}
  \centering
  \begin{tabular}{l@{\hspace{2.5em}}r@{\hspace{1em}}c@{\hspace{1em}}c@{\hspace{1em}}v@{\hspace{1em}}v@{\hspace{1em}}fc@{\hspace{4em}}r@{\hspace{1em}}c@{\hspace{1em}}c@{\hspace{1em}}v@{\hspace{1em}}v@{\hspace{1em}}f}
    \hline
    \hline \\[-1em]
    & \multicolumn{6}{c}{$R=U_4$, \ $U_4^{\rm AF}$} & & \multicolumn{6}{c}{$R=\xi/L$, \ $\xi^{\rm AF}/L$}\\
    \cline{2-7} \cline{9-14}\\ [-0.8em]
        & $L_{\rm min}$ & $m_{\rm max}$ & $k_{\rm max}$ & \multicolumn{1}{c}{$R^*$} & \multicolumn{1}{c}{$T_c$} & \multicolumn{1}{c}{$\chi^2/\text{DOF}$} & & $L_{\rm min}$ & $m_{\rm max}$ & $k_{\rm max}$ & \multicolumn{1}{c}{$R^*$} & \multicolumn{1}{c}{$T_c$} & \multicolumn{1}{c}{$\chi^2/\text{DOF}$} \\
    \hline \\[-1em]
        & $8$  & $1$ &                & 0.4984(1) & 1.2571(1)  & 310/48 && $8$  & $1$ &                & 0.6199(4) & 1.26094(7)& 101/48 \\
        & $12$ & $1$ &                & 0.4960(2) & 1.2583(1)  & 96/35  && $12$ & $1$ &                & 0.6228(6) & 1.26058(9)& 49/35 \\
        & $16$ & $1$ &                & 0.4940(4) & 1.2592(2)  & 44/22  && $16$ & $1$ &                & 0.623(1)  & 1.2606(2) & 22/22 \\
        & $8$  & $2$ &                & 0.4986(2) & 1.2575(1)  & 190/47 && $8$  & $1$ &$0^*$           & 0.636(3)  & 1.2599(2) & 64/47 \\
   FM   & $12$ & $2$ &                & 0.4965(2) & 1.2583(2)  & 40/34  &&      &     &                &           &           &\\
        & $16$ & $2$ &                & 0.4951(4) & 1.2589(2)  & 10/21  &&      &     &                &           &           &\\
        & $8$  & $2$ & $0^{\phantom{*}}$ & 0.4912(6) & 1.2598(2)  & 30/45  &&      &     &                &           &           &\\
        & $8$  & $2$ & $0^*$          & 0.486(1)  & 1.2605(3)  & 30/46  &&      &     &                &           &           &\\[0.2em]
    \hline \\[-0.8em]
        & $8$  & $1$ &                 & 0.4965(2) & 1.3752(1) & 156/43 && $8$  & $1$ &                & 0.5843(5) & 1.38212(8)& 1067/43\\
        & $12$ & $1$ &                 & 0.4949(2) & 1.3760(2) & 64/32  && $12$ & $1$ &                & 0.5986(6) & 1.3801(1) & 122/32 \\
        & $16$ & $1$ &                 & 0.4940(4) & 1.3764(2) & 44/21  && $16$ & $1$ &                & 0.607(1)  & 1.3791(2) & 19/21\\
        & $8$  & $2$ &                 & 0.4966(2) & 1.3754(1) & 91/42  && $8$  & $1$ & $0^{\phantom{*}}$ & 0.68(7)   & 1.375(2)  & 52/41\\
    AF  & $12$ & $2$ &                 & 0.4953(2) & 1.3760(2) & 27/31  && $8$  & $1$ & $0^*$          & 0.672(3)  & 1.3756(2) & 52/42 \\
        & $16$ & $2$ &                 & 0.4949(4) & 1.3762(2) & 17/20  &&      &     &                &           &           &      \\
        & $8$  & $2$ & $0^{\phantom{*}}$  & 0.4947(3) & 1.3762(2) & 31/40  &&      &     &                &           &           &      \\
        & $8$  & $2$ & $0^*$           & 0.489(1) & 1.3773(3)  & 34/41  &&      &     &                &           &           &      \\
    \hline
    \hline
  \end{tabular}
  \caption{Results of fits to Eq.~(\ref{FSSAnsatz}). An absent $k_{\rm max}$ indicates a fit without including scaling corrections. In the fits indicated with $^*$ we fix $\omega=0.8$.}
  \label{tab:fits}
\end{table*}

We have investigated the critical behavior of the model at a finite temperature for vanishing coupling constants $J_2=J_3=0$, considering a FM and AF $J_1$ interaction. Here, we fix $|J_1|=1$.
We have simulated the model by means of the looper code \cite{TK-01,*looperweb} of the ALPS library \cite{ALPS,ALPS2,*alpsweb}, for lattice sizes $L=8$, $12$, $16$, $24$, for a total of $2L^3$ lattice sites, and in an interval around the critical temperature.
To compute the critical temperature we have performed a finite-size scaling \cite{Privman-90,CPV-14} analysis of two renormalization-group invariant quantities. For the FM case we study the Binder ratio $U_4$ and the
ratio $\xi/L$ of the second-moment correlation length $\xi$ over the lattice size $L$. The Binder ratio is defined as $U_4\equiv \<M^2\>/\<M^4\>$, where $M$ is the total magnetization of the system.
The finite-size correlation length $\xi$ is defined in terms of the local magnetization $\hat{S}^z_{i}$.
In the AF case, the order parameter is the staggered magnetization. Accordingly, we have analyzed the staggered Binder ratio $U_4^{\rm AF}\equiv \<M_s^2\>/\<M_s^4\>$, where $M_s\equiv \sum_i \epsilon_i\hat{S}^z_{i}$ is the staggered magnetization, and $\epsilon_i= 1$ ($\epsilon_i=-1$) when the lattice site $i$ belongs to the $A$ ($B$) sublattice. As for the correlation length ratio, we have analyzed the quantity $\xi^{\rm AF}/L$, where the antiferromagnetic second-moment correlation length $\xi^{\rm AF}$ is defined in terms of the local staggered magnetization $\epsilon_i\hat{S}^z_{i}$. A discussion on the definition of a finite-size second-moment correlation length, in terms of a local order parameter, can be found in Appendix A of Ref.~\cite{PTHAH-14}.
In Fig.~\ref{fig:rg} we show the QMC estimates for the renormalization-group invariant observables considered.

Following Refs.~\cite{HPTPV-08,PTPV-09}, to analyze a renormalization-group invariant quantity $R$, we expand the corresponding scaling function and its leading scaling correction in a Taylor series around the critical temperature. To illustrate the procedure, we first consider the FM case and fit the Binder ratio $R=U_4$ to
\begin{equation}
    R(T,L)=R^* +\sum_{m=1}^{m_{\rm max}} a_m(T-T_c)^mL^{m/\nu},
  \label{FSSAnsatz_nocorr}
\end{equation}
where $T_c$ is the critical temperature and $R^*$ is the universal value of $R$ at the critical point. Since the phase transition belongs to the classical three-dimensional Heisenberg universality class, here and in the following we fix the exponent $\nu$ to the corresponding value for such universality class $\nu=0.7112(5)$ \cite{CHPRV-02}.
In Eq.~(\ref{FSSAnsatz_nocorr}) we neglect scaling corrections. To monitor their influence, we have systematically disregarded the smallest lattice sizes.
In Table \ref{tab:fits} we report fit results as a function of $m_{\rm max}$ and the minimum lattice size taken into account $L_{\rm min}$.
For a given value of $L_{\rm min}$ we observe a significant drop in the value of $\chi^2/\text{DOF}$ (DOF denotes the degrees of freedom) when increasing $m_{\rm max}$ from $m_{\rm max}=1$ to $m_{\rm max}=2$, whereas fits for $m_{\rm max}=3$ (not reported here) give a negligible improvement of $\chi^2/\text{DOF}$.
This indicates that a second-order Taylor expansion of $R=U_4$ suitably describes the data, whereas a linear approximation is not sufficient. For fixed $m_{\rm max}=2$, on increasing $L_{\rm min}$ the value of $\chi^2/\text{DOF}$ decreases and we obtain a good value for $L_{\rm min}=12$, $16$. However, we also observe a systematic drift of the fitted values, which is larger than the statistical error bars.
This clearly indicates that scaling corrections are relevant.
To test their influence on the final results, we include them into the analysis, replacing Eq.~(\ref{FSSAnsatz_nocorr}) with
\begin{equation}
  \begin{split}
    R(T,L)=R^* &+\sum_{m=1}^{m_{\rm max}} a_m(T-T_c)^mL^{m/\nu} \\
    &+ L^{-\omega}\sum_{k=0}^{k_{\rm max}}b_k(T-T_c)^kL^{k/\nu},
  \end{split}
  \label{FSSAnsatz}
\end{equation}
where $k_{\rm max}$ is the Taylor expansion order of the correction to scaling term.
Within the range and precision of our QMC data, fits of $R=U_4$ with $m_{\rm max}=2$ allow to include corrections-to-scaling with $k_{\rm max}=0$, i.e., to the leading order only, providing a suitable approximation of the scaling function and allowing to extract consistent results. As a further check, we have repeated the fits to Eq.~(\ref{FSSAnsatz}) fixing the value of the correction-to-scaling exponent $\omega=0.8$, as expected for the three-dimensional $O(3)$ universality class \cite{CHPRV-02}.
A similar analysis has been done with the renormalization-group invariant ratio $R=\xi/L$. In this case we find that a linear approximation $m_{\rm max}=1$ is sufficient to fit the data. In Table \ref{tab:fits} we report the results of our fit.
Along these lines we have also analyzed the AF cases. As for the FM case, we have found that, within the range of our data and their precision, suitable fits of the Binder ratio $U_4^{\rm AF}$ require $m_{\rm max}=2$, whereas for $\xi^{\rm AF}/L$ a linear approximation $m_{\rm max}=1$ is sufficient. Corresponding fits are reported in Table \ref{tab:fits}.
By judging conservatively the fit results, we extract the estimates of the critical temperatures for the FM and AF models reported in Table \ref{tab:TCTN}.

Fits in Table \ref{tab:fits} allow also to determine the universal values $R^*$ of the renormalization-group invariant quantities at criticality. As expected by universality, they are the same for the FM and AF model. Inspecting the fit results of Table \ref{tab:fits} we estimate
\begin{align}
  &U_4^* = U_4^{\rm AF *} = 0.491(5), \label{binder_star}\\
  &(\xi/L)^* = (\xi^{\rm AF}/L)^* = 0.62(2), \label{xil_star}
\end{align}

\end{document}